\documentstyle[12pt,psfig]{article}
\newcommand{\mysection}{\setcounter{equation}{0}\section}

\def\beq{\begin{equation}}
\def\eeq{\end{equation}}
\def\beqa{\begin{eqnarray}}
\def\eeqa{\end{eqnarray}}

\begin{document}

\begin {flushright}
EDINBURGH 97/3\\
ITP-SB-97-24
\end {flushright} 
\vspace{3mm}
\begin{center}
{\Large \bf Resummation 
for QCD Hard Scattering} 
\end{center}
\vspace{2mm}
\begin{center}
Nikolaos Kidonakis\\
\vspace{2mm}
{\it Department of Physics and Astronomy\\
University of Edinburgh\\
Edinburgh EH9 3JZ, Scotland} \\
\vspace{4mm}
George Sterman \\
\vspace{2mm}
{\it Institute for Theoretical Physics\\
State University of New York at Stony Brook\\
Stony Brook, NY 11794-3840, USA} \\  
\vspace{2mm}
May 1997
\end{center}

\begin{abstract}
We resum distributions that are singular
at partonic threshold (the elastic limit) in
heavy quark production, in terms of
logarithmic behavior in moment space.
The method may be applied to a variety
of cross sections sensitive to the edge of phase
space, including transverse momentum distributions.
Beyond leading logarithm, dependence 
on the moment variable is controlled by
a matrix renormalization group equation,
reflecting the evolution of composite
operators that represent the color
structure of the underlying hard scattering.
At next-to-leading logarithmic accuracy, 
these evolution equations may be diagonalized,
and moment dependence
in the cross section is a sum of exponentials.  
Beyond next-to-leading logarithm, 
resummation involves matrix-ordering.
We give a detailed analysis for the 
case of heavy quark production by light quark annihilation
and gluon fusion.
\end{abstract}

\pagebreak

\mysection{Introduction}

The factorization of cross sections into long- and short-distance
functions gives QCD predictive power for a wide variety of inclusive,
hard-scattering cross sections in hadronic collisions.
There has been great progress in recent years in the calculation
of perturbative corrections at next-to-leading order (NLO) for such processes,
including Drell-Yan \cite{dy1loop}, direct photon \cite{dgamma1loop},
heavy quark \cite{heavycalcs} and jet production \cite{jet1loop}. 
Selected exact results at two or even more loops are also available, especially
for ${\rm e}^+{\rm e}^-$ annihilation  \cite{epemho} and
deeply inelastic scattering \cite{2loop}.
The exploration of quantum chromodynamics, electroweak
theory, and searches for extensions of the standard model, 
however, require information on yet higher orders.
We need to develop methods to decide when higher-order perturbative
terms are negligible and, when they are not, to 
estimate their importance.  
This motivates the systematic study 
of pertubation theory at arbitrary order.
One of the elements in this program is the resummation 
of singular distributions in factorized inclusive cross sections.
This resummation is most conveniently carried out in
moment space.  Closely related resummations organize 
large logarithmic corrections near zero transverse momentum
for heavy systems such as Drell-Yan or heavy quark pairs.

In this paper we shall give details of this resummation 
in hard-scattering cross sections 
induced by QCD itself, particularly heavy quark production, 
extending results for electroweak-induced hard scattering
cross sections, such as the Drell-Yan production of vector bosons \cite{oldDY,CT1,CT2}.
Our starting point will be the factorization properties of
these cross sections near threshold, and we shall make
strong use of the close relation between factorization and 
resummation \cite{CLS}.

The resummations that we consider apply to  
 processes of the form
\beq
h_1(p_1)+h_2(p_2) \rightarrow T(q,y,\chi)+X\, ,
\eeq
where $T(q,y,\chi)$ represents a 
system in the hadronic final state, produced
with large
invariant mass $\sqrt{q^2}=Q$ and (collective) rapidity $y$, 
\beq
y={1\over 2}\ln\left({q\cdot p_2\over q\cdot p_1}\right)\, .
\label{repidef}
\eeq 
Examples include the production of heavy quark pairs and
high-$p_T$ jets. 
 The additional variable $\chi$
represents the internal structure of the final  state, for example
the angle $\theta$ between the  quark, or jet, direction
in the final state and the beam axis.
 Such cross sections may be written in factorized form as
\beqa
\frac{d\sigma_{h_1\; h_2}}{dQ^2 \; d\chi \; dy}&=&\sum_{ab} \; 
\int {dx_a} {dx_b}\,  \phi_{a/h_1}(x_a,\mu^2) 
\phi_{b/h_2}(x_b,\mu^2)
\nonumber \\ &&
\times\ H_{ab}\left(\frac{Q^2}{x_a x_b S},y-y_p,\chi,{Q\over \mu},Q,
\alpha_s(\mu^2)\right) ,
\label{factth}
\eeqa
where $S$ is the center-of-mass (c.m.) energy squared of the hadrons,  and where
the $\phi$'s are parton distributions defined in an appropriate
scheme, and evaluated at factorization scale $\mu$.  The function
$H_{ab}$ is the hard-scattering partonic cross section (the ``hard part")
that describes the production of the final-state system $T$ 
in the collision of partons $a$ and $b$, which carry
momentum fractions $x_a$ and  $x_b$
of hadrons $h_1$ and $h_2$, respectively.  Rapidity-dependence
in $H_{ab}$ is through the combination $y-y_p$, where $y_p$ is
the partonic rapidity in the hadronic c.m. frame,
\beq
y_p={1\over 2}\ln\left({x_a\over x_b}\right)\, .
\label{parrapidef}
\eeq
The hard parts $H_{ab}$ in Eq.\ (\ref{factth})
are calculable in perturbation
theory.  In general, they are not smooth functions of their
arguments, but combinations of functions and
singular distributions.  The reason for this somewhat 
complicated structure is  that they summarize 
cancellations between cross sections with gluon emission and with
virtual gluon corrections.  This cancellation (in QCD as in QED) 
is only manifest at the level of 
sufficiently inclusive cross sections.  For factorized
cross sections such as Eq.\ (\ref{factth}), 
this requirement manifests itself in the occurrence of
 (``plus")
distributions, which are singular  for $z=1$, where
\begin{equation}
z={Q^2\over s},
\label{def}
\end{equation}
with $s=x_ax_bS$ the invariant mass squared
of the partons that initiate the hard scattering.  
We shall refer to $z=1$ as  ``partonic threshold" 
\footnote{We emphasize that by partonic threshold, we
refer to the c.m.\ total energy of the incoming partons
for a fixed final state; heavy
quarks, for example, are not necessarily produced at rest.},
or sometimes as the ``elastic limit". 

Singular distributions 
are most easily organized in terms of moments 
with respect to $\tau=Q^2/S$.  As we review below,
for the Drell-Yan and related cross sections
logarithmic dependence on the moment variable 
exponentiates \cite{oldDY}-\cite{veryoldDY}.  This moment must
be inverted to derive the physical cross section.
There are a number of proposals on the best 
way to carry out this transformation, 
or otherwise use the resummation, which have been
put forward in the context of
Drell-Yan \cite{Appeletal,CSt,AC} and
heavy quark production \cite{LSN1,bc,cat,HERAB,nkjsrv,jadward}.
Our analysis here will not deal directly with
this question; our results will remain in moment
space.  In this paper, we shall be concerned with
the extension of resummation to cross sections that are
initiated by QCD itself.  Our explicit applications
will be to resummation at partonic threshold for heavy quark
production, but the method is much more general.  
A straightforward application of the method, which involves
only the replacement of the moments by Fourier transformation
to impact parameter space, is to the overall transverse momentum
of the heavy quark pair, again extending results for
Drell-Yan pairs \cite{DYqt}, and hadrons in ${\rm e}^+{\rm e}^-$ 
annihilation \cite{bbj}.

It has been recognized
for some time that leading logarithms in moment (or impact parameter)
space
exponentiate for heavy quark production \cite{LSN1,bc,cat} in the
same manner as for Drell-Yan,
reflecting the factorizability of leading singular
distributions from the hard subprocess \cite{oldDY}.
Beyond the level of leading logarithms, however, singular distributions
in QCD-induced cross sections depend
coherently on color exchange in the underlying hard scattering.
The interplay between the color structures of short and long
distance interactions will
be a unifying thread of the discussion that follows.

The next section starts with a brief review of the 
factorization of the inclusive Drell Yan cross sections
near partonic threshold.   We shall describe the resummation
of the singular distributions in these cross sections 
starting from their  factorization properties.  We also 
observe the field-theoretic content of the functions
into which the cross section factorizes,
relating them to matrix elements of
fields and of Wilson lines \cite{eikrenorm,BottsSt,GK,KK,KoSt}.
We then extend the factorization, and hence resummation, formalism to
QCD-induced inclusive cross sections.  The possibility for, and the broad
outlines of, such an extension has recently been discussed \cite{CLS}; here,
the practical formalism, and its field theoretic basis, are
discussed in some detail, expanding on a
summary given in Ref.\ \cite{NKGS} (see also \cite{Thesis}). 
To organize sensitivity to color exchange at
short distances
beyond leading order in $\ln(1-z)$ (or in logarithms of the
corresponding moment variable), 
we derive a matrix evolution equation
that controls all logarithmic enhancements at
threshold.  At next-to-leading logarithm (NLL)
in the exponent, the 
matrix equation may be diagonalized in a manner
reminiscent of leading order evolution in deeply inelastic scattering.
At NLL, the
cross section is a sum of exponentials,
each representing an eigenvector of the anomalous dimension
matrix.  Beyond this level of accuracy, however, the resummation
can only be carried out in terms of ordered exponentials.
Resummation at NLL requires the computation of
the anomalous dimension matrix for each partonic
subprocess, such as light quark annihilation into 
heavy quarks or gluon fusion into heavy quarks.

The third section describes
the calculation of the anomalous dimension matrix for 
light quark annihilation into heavy quarks, supplementing the
brief description of Ref.\ \cite{NKGS}.  Section 4 extends these
results to the very important case of gluon fusion. The
paper concludes with a summary and the outlook for 
further progress.

\mysection{Factorization and Resummation}

\subsection{Resummation for Color Singlet Hard Scattering}

The resummation of singular distributions for the Drell-Yan,
and related electroweak-induced hard scattering processes has
been understood for some time \cite{oldDY,CT1,CT2}.  
In this subsection, we review the basic results,
and elaborate somewhat on the operators in QCD that 
play a role in the underlying factorization.
For our purposes, it will be useful to generalize the standard
results slightly, to the production of an arbitrary color-singlet
system, denoted $B$ below, of invariant mass 
squared $Q^2$.  We assume in particular
that $B$ may be produced by gluon fusion as well as quark annihilation.
An important example is Higgs production \cite{Higgs}.
For the purposes of the following discussion, however, we shall refer to
this class of production processes as Drell-Yan cross sections.

Resummation is most readily 
formulated in terms of moments of the 
inclusive Drell-Yan cross section $d\sigma/dQ^2$,
with respect to the variable $\tau=Q^2/S$,
\beqa
\int_0^1 d\tau \tau^{N-1} {1\over \sigma_0}
{d\sigma_{h_1h_2\rightarrow B} \over d Q^2}(\tau, Q^2) 
 &=&   \sum_{i=q,{\bar q},g}\; e_f^2\; {\tilde \phi}_{i/h_1}(N,\mu^2) \, 
{\tilde \phi}_{{\bar i}/h_2}(N,\mu^2) \cr
 &\ & \times {\tilde \omega}_{i{\bar i}\rightarrow B}
\left(N, Q/\mu,\alpha_s(\mu^2)\right)+{\cal O}(1/N)\, . \nonumber
\\
&\ & 
\label{dymoment}
\eeqa
Here, $e_f$ is the fractional
quark charge, $\sigma_0\ (=4\pi\alpha^2/(9Q^2S)$ for the standard Drell-Yan cross
section) is a Born cross section,
and the moment functions $\tilde\phi$ are
\beq
{\tilde \phi}(N,\mu^2)=\int_0^1dx\; x^{N-1}\phi(x,\mu^2)\, ,
\eeq
and similarly for the hard-scattering function 
$\omega_{i{\bar i}\rightarrow B}(z,Q/\mu,\alpha_s(\mu^2))$
in terms
of the variable $z$.
In Eq.\ (\ref{dymoment}) we have neglected hard-scattering functions
associated with the gluon-quark combination, with two quarks or antiquarks,
and with quarks and antiquarks of different flavors, since 
these do not give rise to
singular distributions at $z=1$.  
All such corrections, which are smooth at $z=1$, contribute
to the moments only by corrections that vanish as powers
of $N$ for $N\rightarrow \infty$, as indicated. 
Singular distributions, however, give rise to 
logarithmic $N$-dependence, as in
\beq
\int_0^1 dz\; z^N\; 
\left [ {\ln^m(1-z)\over 1-z} \right ]_+
 = {(-1)^{m+1} \over m+1} \ln^{m+1}N+\dots
\label{logNfromsd}
\eeq
 Thus,
we should think of the large-$N$ behavior as a 
diagnostic for singular distributions in $1-z$.
Physical cross sections, of course, always involve
the inverse transform of the moments.
  
The functions $\omega_{i{\bar i}\rightarrow B}(z,Q/\mu,\alpha_s(\mu^2))$
whose moments appear in Eq.\ (\ref{dymoment})
are normalized to
\beq
\omega_{i{\bar i}\rightarrow B}=\delta(1-z)+{\cal O}(\alpha_s)\, .
\label{omeganorm}
\eeq
We shall see that the 
logarithmic $N$-dependence of moments of $\omega$ exponentiates
into a useful form.  

Before sketching the 
physical basis and derivation of this result, we observe
that, as shown in \cite{LaenenSterman}, the same functions 
$\omega_{i{\bar i}\rightarrow B}$ as in Eq.\ (\ref{dymoment})
contain the singular distributions at $z=1$ for
fixed rapidity,
\beqa
{d\sigma_{h_1h_2\rightarrow B}\over dQ^2 dy}
&=&
\sum_{i=q,{\bar q},g}\, 
\sigma_0\, \int_\tau^1 dz\int {dx_a\over x_a} {dx_b\over x_b}\, 
\phi_{i/h_1}(x_a,\mu^2)\; \phi_{{\bar i}/h_2}(x_b,\mu^2)
 \cr
&\ & \ \times\; \delta\left (z-{Q^2\over x_ax_bS}\right)\; 
\delta\left( y-{1\over2}\ln{x_a\over x_b}   \right)
\omega_{i{\bar i}\rightarrow B}(z, Q/\mu,\alpha_s(\mu^2))\, .
\cr
&\ &
\label{rapomega}
\eeqa
Thus, we may compute singular threshold corrections
for the cross section at fixed rapidity from 
the inclusive cross section.

The resummation of singular distributions in $1-z$ may be carried out
in a number of  ways, 
but all are based on the observation that the perturbative function $\omega_{i{\bar i}}$
may be computed in dimensionally regulated perturbation theory by choosing the 
external hadrons to be {\it partons}.
For us the most informative approach is based on
the ability to reexpress moments of the inclusive Drell-Yan cross section
Eq.\ (\ref{dymoment}), with, for instance, $h_1=q$, $h_2={\bar q}$ \cite{oldDY,CT1}.
In this form, all logarithmic $N$ behavior is factorized
into alternate parton distributions $\psi$, along with a 
``soft-parton" function $U$,
which describes the exchange and emission of soft gluons by the
annihilating quark pair \cite{oldDY},
\begin{eqnarray}
\int_0^1 d\tau \tau^{N-1} {1\over\sigma_0}
{d\sigma_{i{\bar i}\rightarrow B} \over d Q^2}(\tau, Q^2,\epsilon) 
&=& 
\bigg | H \left({Q\over\mu},\alpha_s(\mu^2)\right)\bigg|^2
{\tilde U}^{(i)} \biggl ( {Q\over N\mu },\alpha_s(\mu^2) \biggr ) \cr
&\ & \quad \quad \times
{\tilde\psi}_{i/i}\left ( N,{Q\over \mu },\epsilon \right )
\;
{\tilde\psi}_{{\bar i}/{\bar i}}\left (N,{Q\over \mu },\epsilon \right ) \cr
&\ &\hbox{\hskip 0.5 true in}
+{\cal O}(1/N)\, .
\label{qqgammastarfact}
\end{eqnarray}
This factorization is illustrated in Fig.\ 1.
$H$ is an infrared-safe function that is independent of $N$. 
Let us discuss the remaining functions, which absorb the $N$-dependence,
in turn.    
\begin{figure}
\centerline{
\psfig{file=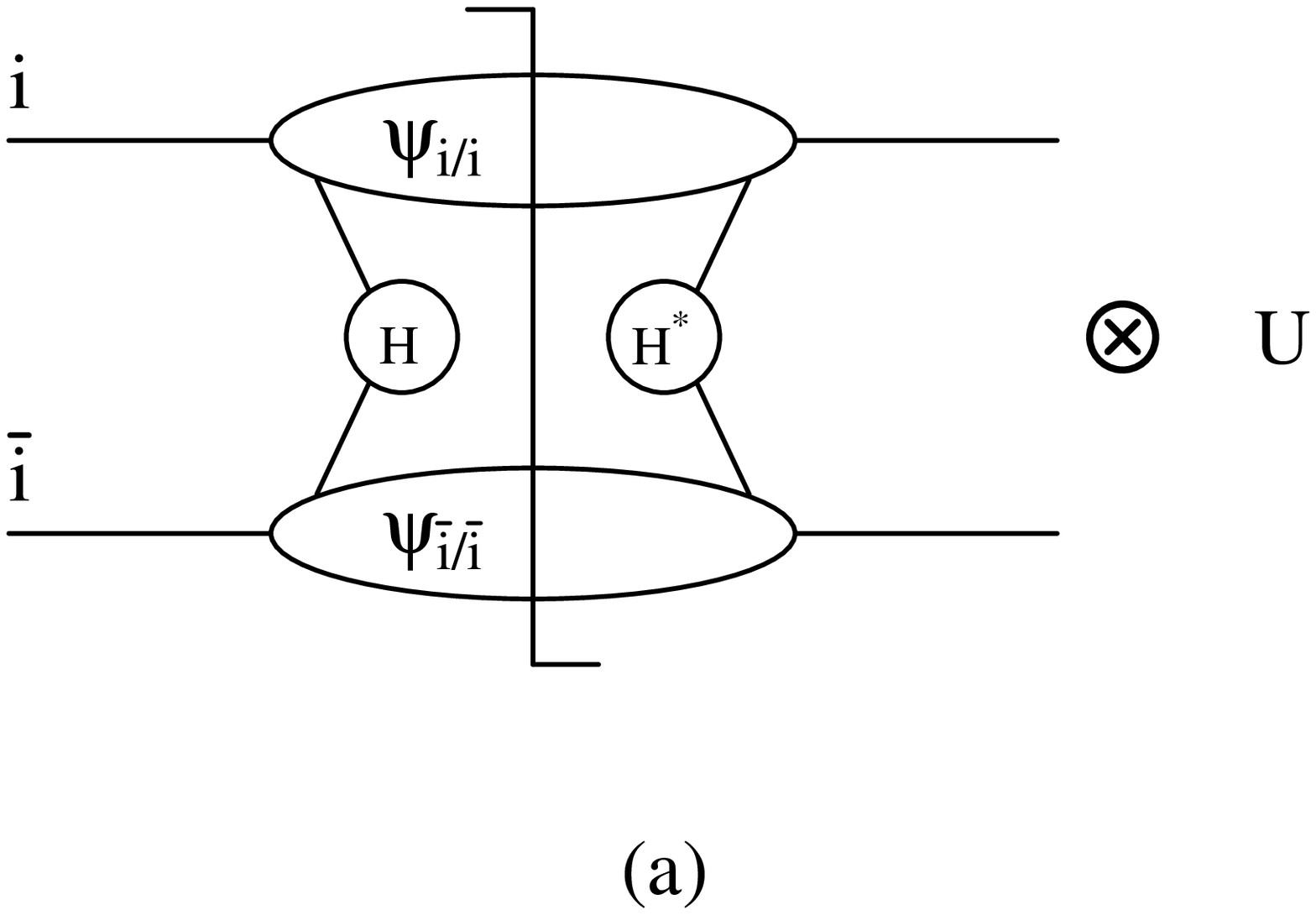,height=2.5in,width=4.05in,clip=}}
\vspace{5mm} 
\centerline{
\psfig{file=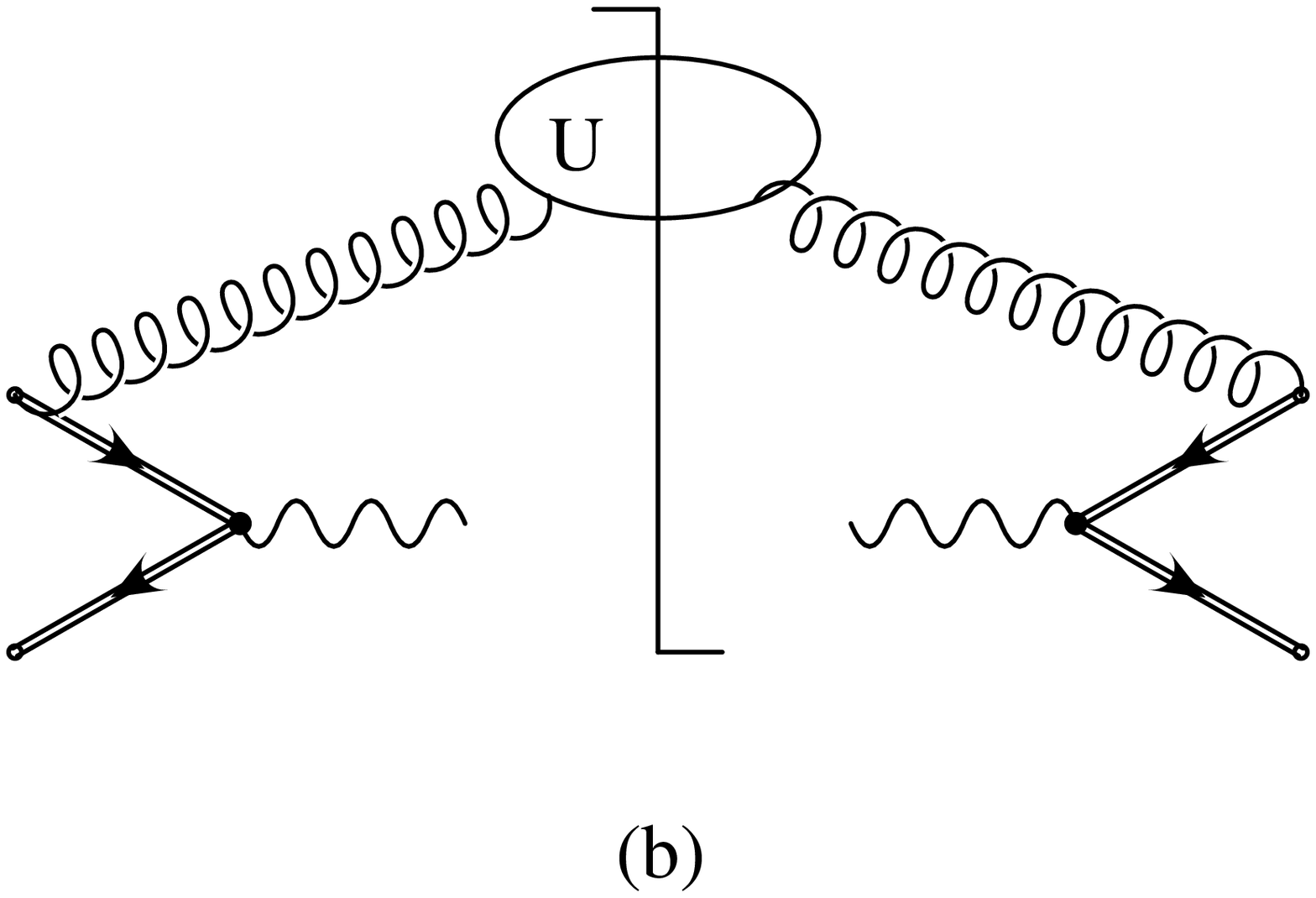,height=2.5in,width=4.05in,clip=}}
{Fig. 1. (a) Factorization for the Drell-Yan process near partonic threshold,
Eq.\ (\ref{qqgammastarfact}).
(b) The soft-gluon function $U$. The double lines
represent ordered exponentials in Eq.\ (\ref{WDY}).}
\label{fig. 1}
\end{figure}

The modified parton distributions $\psi$ in Eq.\ (\ref{qqgammastarfact}) are 
conveniently defined in the partonic center of mass system
at fixed {\it energy}, rather
than a light-like momentum fraction.
For $i=q$, for example, we have\footnote{This form includes an explicit
spin average.  Single-particle states are normalized
by $\langle 0|q(0)|p,s\rangle=u(p,s)$.  
The factor of $1/2^{3/2}$ then normalizes the lowest
order to $\delta(1-x)$.}
\begin{equation}
\psi_{q/q}(x,2p_0/\mu,\epsilon)
=
{1\over 2\pi 2^{3/2}}
\int_{-\infty}^\infty dy_0\ {\rm e}^{-ixp_0y_0}
\langle q(p)|\bar{q}(y_0,\vec{0})\; {1\over 2}n\cdot \gamma\; q(0)|q(p)\rangle\, ,
\label{psidef}
\end{equation}
where the matrix element is evaluated in $A_0=0$ gauge
in the partonic c.m.\ frame.  The vector $n$ is lightlike in
the opposite direction from $p^\mu$, so that for $\vec{p}$ in
the $\pm 3$ direction, $n\cdot\gamma=\gamma^\mp$.  
$\psi_{{\bar q}/{\bar q}}$ and $\psi_{g/g}$ may
be constructed similarly.
Defined in this fashion, $\psi$ 
absorbs all the collinear divergences of the Drell-Yan
cross section in the $\tau\rightarrow 1$ limit \cite{oldDY}.
Its moments may thus
be factorized into 
a product of moments of the corresponding
light-cone parton distribution $\phi$, defined in any scheme,  times
an infrared safe function.

The function $U^{(i)}$ represents the coupling of soft gluons
to the annihilating partons.  So long as these soft gluons are
not collinear to the active partons,
this coupling is well approximated by ordered
exponentials, or Wilson lines, which we denote as
\begin{equation}
\Phi_\beta(\lambda_2,\lambda_1;x)
=
P\; \exp \left (-ig\int_{\lambda_1}^{\lambda_2} d\lambda' \beta\cdot A(\lambda'\beta+x)
\right )\, ,
\label{phidef}
\end{equation}
with $A$ the gauge field in the appropriate representation
of the gauge group, and $\beta$ a velocity four-vector,
in the direction of the particle whose interactions
with soft gluons are being approximated.  In the case of 
a quark, the gauge fields are in the fundamental representation,
for a gluon in the adjoint.  The operator $P$ orders group 
products in the same sense as ordering in the integration
variable $\lambda$, with the $A$'s with lower values
of $\lambda$ to the right.  In this notation, 
$\Phi_\beta(\lambda_2,\lambda_1;x)$
represents (for example) a quark that propagates from $x+\lambda_1\beta$
to $x+\lambda_2\beta$, while $\Phi_\beta(\lambda_1,\lambda_2;x)$
represents an antiquark with the same propagation.  The construction
of $U^{(i)}$ from the $\Phi$'s is actually a two-step process, which we
now describe.

Starting with the Wilson lines, we first construct
a (dimensionless) ``eikonal singlet cross section"
\beqa
\sigma^{({\rm eik})}_{\rm DY}\left((1-w)Q/\mu,\alpha_s(\mu^2),\epsilon\right)
&=&
{Q\over 2\pi}\; \int d x_0\; {\rm e}^{i(1-w)Qx_0}\; 
\langle 0| {\bar T}\left[W^\dagger_{b_1b_2}(x_0,\vec{0}) \right]\cr
&\ & \quad \quad \times\,  
T\left[W_{b_2b_1}(0)\right]|0\rangle\, ,
\label{udef}
\eeqa
where
$W(x)$ is the product of Wilson lines representing the pair,
coupled by a color-singlet vertex at the point $x$,
\begin{equation}
W_{b_2,b_1}(x)
=
\delta_{a_1a_2}\; 
\Phi_{\beta_2}(-\infty,0;x)_{b_2a_2}\Phi_{\beta_1}(0,-\infty;x)_{a_1b_1}\, .
\label{WDY}
\end{equation}
In Eq.\ (\ref{udef}), $T$ represents time-ordering, and $\bar T$ anti-time ordering.
When dealing with QCD hard-scattering below, we shall generalize
this product, by incorporating nontrivial color structure at the vertex between
the lines.  The quantity $(1-w)Q$ is the energy emitted by the Wilson
lines \cite{KoSt,KS}, which is the relevant measure of phase space near
partonic threshold in the center of mass frame 
of the colliding partons \cite{oldDY}.  In Eq.\ (\ref{udef}),
an appropriate normalization is understood, so that at lowest order
$\sigma^{\rm (eik)}_{\rm DY}=\delta(1-w)$.

The eikonal cross section represents the emission of soft gluons
from the annihilating pair.  The scale invariance of the Wilson
lines implies, however, that it contains ultraviolet, in addition
to infrared and collinear, divergences, corresponding to the
renormalization of the composite operator introduced by the
vertex in Eq.\ (\ref{WDY}) where the eikonal lines join \cite{eikrenorm,BottsSt,GK,KK}.
A natural ``scheme" for this renormalization is specified 
 by demanding that the integral of $\sigma^{\rm (eik)}_{\rm DY}(1-w)$
from $w=0$ to $w=1$ be normalized to  unity,
\beq
\int_0^1 dw\; \sigma^{\rm (eik)}_{\rm DY}\left((1-w)Q/\mu,\alpha_s(\mu^2),\epsilon\right)
=1\, .
\label{sigeikdynorm}
\eeq
That is, we demand that all corrections to $\sigma^{\rm (eik)}_{\rm DY}$ 
vanish for the fully inclusive cross section.
With this normalization, beyond lowest order $\sigma^{\rm (eik)}_{\rm DY}$ is a sum of
``plus" distributions in $1-w$,
\beqa
\sigma^{\rm (eik)}_{\rm DY}\left((1-w)Q/\mu,\alpha_s(\mu^2),\epsilon\right)
&=& \delta(1-w) \nonumber \\
&\ & \hskip -0.1 true in + 
\sum_{n\ge 0} s_n\left(Q/\mu,\alpha_s(\mu^2),\epsilon\right)\; 
\left [{\ln^n(1-w) \over 1-w}\right]_+\, ,\nonumber
\\
&\ & \label{sigeikdyasplus}
\eeqa
where the coefficients $s_n$ are power series in the coupling.

Even after renormalization,
$\sigma^{\rm (eik)}_{\rm DY}$ includes the couplings of 
fast-moving gluons
that are collinear to the incoming Wilson lines.  Such gluons, whether virtual
or real,  are already included in the distributions $\psi$, 
and their contributions should be removed from the soft
gluon function, $U^{(i)}$, to avoid double counting.  This is possible because noncollinear
soft gluons factor from collinear gluons through the 
use of Ward identities \cite{fact}.  
Physically, soft radiation cannot
distinguish between physical and eikonal light-like sources.  
Thus, moments of $\sigma^{\rm eik}_{\rm DY}$ 
are factorizable in the same manner as the 
quark-antiquark cross section cross section, Eq.\ (\ref{qqgammastarfact}), into
moments of the (identical) soft-parton function $U$ times 
moments of ``jet" functions ${\tilde j}_{1,2}$, analogous to the $\psi$'s,
Eq.\ (\ref{psidef}),
but with the incoming partons represented by 
products of eikonal lines
(in $A_0=0$ gauge) as
\beqa
j_1\left({(1-w_1)Q\over \mu},\alpha_s(\mu^2),\epsilon\right) 
&=&
{Q\over 2\pi}
\int_{-\infty}^\infty dy_0\ {\rm e}^{-i(1-w_1)Qy_0}\cr
&\ & \ \times
\langle 0|\; {\rm Tr}\bigg\{\; {\bar T}[\Phi_{\beta_1}^\dagger(0,-\infty;y)]\cr
&\ & \ \ \times
 T[\Phi_{\beta_1}(0,-\infty;0)]\; \bigg\}\; |0\rangle\, ,
\label{eikjdef}
\eeqa
with $y^\nu=(y_0,\vec{0})$ a vector at the spatial origin.
The ``antiquark" distribution $j_2$ is defined analogously,
but with the Wilson line in the $\beta_2$ direction and ordered
from $0$ to $-\infty$.

In these terms, the eikonal version of 
the moment relation Eq.\ (\ref{qqgammastarfact}) is
\beqa
{\tilde \sigma}^{({\rm eik})}_{\rm DY}\left(Q/N\mu,\alpha_s(\mu^2),\epsilon\right)
&=&
{\tilde U}^{(i)}\left({Q\over N\mu},\alpha_s(\mu^2)\right)\cr
&\ & \quad \times
{\tilde j}_1\left({Q\over N\mu},\alpha_s(\mu^2),\epsilon\right)\; 
{\tilde j}_2\left({Q\over N\mu},\alpha_s(\mu^2),\epsilon\right)\, ,
\cr &\ &
\label{sigeikDY}
\eeqa
which may be used to construct the soft function $U$.
This prescription eliminates diagrams
that have divergences from gluons collinear to the external eikonal lines of $U$,
a fact which we shall use below.  Because the $j_i$'s absorb all
collinear divergences, $U$ is an infrared safe sum of plus distributions.

Logarithms of $N$ exponentiate in each of the factors of 
Eq.\ (\ref{qqgammastarfact}) \cite{oldDY}.
As emphasized in Ref.\ \cite{CLS}, this is a result of the 
constraint that the product is independent of both the gauge choice
and the factorization scale, although its individual factors are not.
Such combined constraints organize logarithms associated with both
soft and collinear momentum configurations, and are at the basis of
Sudakov resummation \cite{expon}.    
Comparing Eq.\ (\ref{dymoment}) 
with $h_1=i$, 
$h_2={\bar i}$ with
Eq.\  (\ref{qqgammastarfact}), we can
solve for ${\tilde \omega}(N)$, in terms of the calculable $N$-dependence
of $\psi$, $\phi$  and $U^{(i)}$ \cite{oldDY},
\beq
{\tilde \omega}_{i{\bar i}\rightarrow B}(N)=
\left[{{\tilde \psi}_{i/i}(N,1,\epsilon)\over{\tilde \phi}_{i/i}(N,Q^2,\epsilon)}\right]^2\,
\bigg | H \left({Q\over\mu},\alpha_s(\mu^2)\right)\bigg|^2
{\tilde U}^{(i)} \biggl ( {Q\over N\mu },\alpha_s(\mu^2) \biggr )\, ,
\label{omegaqqofN}
\eeq
where we have used $\psi_{i/i}=\psi_{{\bar i}/{\bar i}}$,
and similarly for the $\phi$'s, and have suppressed various arguments.

The result of this procedure applied to the Drell-Yan cross section is
\begin{eqnarray}
{\tilde \omega}_{i{\bar i}\rightarrow B}
(N)&=&A(\alpha_s(Q^2))\; \exp \Bigg [-\int^1_0 dz \frac{z^{N-1}-1}{1-z} 
\nonumber \\ &&
\times \left(\int^{(1-z)^{m_S}}_{(1-z)^2} \frac{d\lambda}{\lambda}
g_1^{(i{\bar i})}[\alpha_s(\lambda Q^2)]
 +g_2^{(i{\bar i})}[\alpha_s((1-z)^{m_s} Q^2)]\right)\Bigg] \, ,
\nonumber\\
&\ &
\label{omegaexp}
\end{eqnarray}
where for DIS ($\overline{\rm MS}$) factorization scheme $m_S=1\, (0)$.
$A(\alpha_s)$ is a constant function of the coupling, and
the $g_{1,2}$ are finite functions of their arguments. 
$g_1^{(ab)}$ and $g_2^{(ab)}$ are universal among hard cross sections and color
structures for given incoming partons $a$ and $b$, but depend on whether these
partons are quarks or gluons.
To reach the accuracy of
next-to-leading logarithms in the exponents, 
we need $g_1$ only to two loops, with leading logarithms coming entirely from
its one-loop approximation,
and $g_2$ only to
a single loop.    More explicitly,
we take \cite{CSt}
\begin{equation}
g_1^{(ab)} = (C_a+C_b)\left ( {\alpha_s\over \pi} 
+\frac{1}{2} K \left({\alpha_s\over \pi}\right)^2\right )\, ,
\label{g1def}
\end{equation}
with $C_i=C_F\ (C_A)$ for an incoming quark (gluon), and
with $K$ given by
\begin{equation}
K= C_A\; \left ( {67\over 18}-{\pi^2\over 6 }\right ) - {5\over 9}n_f\, ,
\label{Kdef}
\end{equation}
where $n_f$ is the number of quark flavors.  
$g_2$ is given for quarks in the DIS scheme by
\begin{equation}
g_2^{(q {\bar q})}=-{3\over 2}C_F\; {\alpha_s\over\pi}\, , 
\end{equation}
and it vanishes in the $\overline {\rm MS}$ scheme.
The resummed $N$-dependence thus depends on the 
factorization scheme, and is quite different between DIS and
${\rm {\overline MS}}$ schemes \cite{CT1,CLS}.  This difference is in principle
compensated for by differences in the $N$-dependence of
the distributions themselves.

The leading logarithms in the exponent of Eq.\ (\ref{omegaexp}) are from
the ratio of $\psi$ to $\phi$ in Eq.\ (\ref{omegaqqofN}).
The function $U^{(i)}$ contributes at next-to-leading logarithm.  To achieve this
accuracy we need only a one-loop calculation, and we find \cite{oldDY}
\beqa
{\tilde U}^{(i)}(N)&=&\exp\left[\int^1_0 dz \frac{z^{N-1}-1}{1-z} 
\nu^{(i)}(\alpha_s((1-z)^2 Q^2))\right] \cr
&=&\exp \left[\int^1_0 dz \frac{z^{N-1}-1}{1-z}
\left(\int^{(1-z)^{m_S}}_{(1-z)^2} \frac{d\lambda}{\lambda} 
\beta(g(\lambda Q^2))\nu^{(i)}[\alpha_s(\lambda Q^2)]\right.\right.
\nonumber \\ &&
\left.\left.\quad \quad +\nu^{(i)}[\alpha_s((1-z)^{m_s} Q^2)]\right)\right]\, ,
\label{Uresum}
\eeqa
with $\nu^{(q,g)}=2C_{F,A}(\alpha_s/\pi)+\dots$  
In the second form, $\beta(g)$ is the QCD
beta function, and we have followed \cite{CT2}
in reexpressing $U^{(i)}$ into
the form of Eq.\ (\ref{omegaexp}).  
In this connection, we note 
that in axial gauges $U^{(i)}$ receives noncancelling contributions only from 
the ``gauge" term $k^{\mu} \, k^{\nu}/(n\cdot k)^2$ in the gluon
propagator.  At least to this order,
it is natural to use Ward identities to factor such terms into the
parton distributions $\psi$.

\subsection{Resummation for QCD Hard Scattering}

Generalizing Eq.\ (\ref{rapomega}),
we consider the production of a pair of heavy quarks
(or jets) at rapidity $y$,
and at scattering angle $\theta$ in the pair center of mass frame, as
\beqa
{d\sigma_{h_1h_2\rightarrow Q{\bar Q}}\over dQ^2 dy d\cos\theta}
&=&
\sum_{f=q,{\bar q},g}\, \int_\tau^1 dz\, 
\int {dx_a\over x_a} {dx_b\over x_b}\, 
\phi_{f/h_1}(x_a,\mu^2)\; \phi_{{\bar f}/h_2}(x_b,\mu^2)
 \cr
&\ & \quad \quad \times\; 
\delta\left (z-{Q^2\over x_ax_bS}\right)\,
\delta\left( y-{1\over2}\ln{x_a\over x_b}   \right)
 \nonumber \\
&\ & \quad \quad \times 
{\hat \sigma}_{f{\bar f}\rightarrow Q{\bar Q}}\left(z, 
\frac{Q}{\mu},Q,\theta,\alpha_s(\mu^2)\right)\cr
&\ &\ \ + \dots \, .\label{rapsigmahat}
\eeqa
Corrections, due to other partonic flavor combinations, are
less singular by a power of $1-z$  as $z\rightarrow 1$ in the hard scattering, and
their moments are suppressed by powers of $1/N$, as in Eq.\ (\ref{dymoment}).
For definiteness, we
shall concentrate on heavy quark production, but our
considerations are more general.
As in the case of Drell-Yan cross sections, the
singular behavior of the
hard-scattering functions ${\hat \sigma}$ 
(the analog of $\omega_{q{\bar q}}$ above)
at fixed rapidity
may be
computed from the inclusive partonic cross section \cite{LaenenSterman}.
Following Ref.\ \cite{CLS}, we begin with the
factorization properties of heavy quark production 
and other QCD hard-scattering processes in
the elastic limit.   We shall see that these depend upon
the directions of the outgoing heavy quarks in the
partonic center of mass frame.  As a result, resummation must
be carried out at fixed directions for the heavy quarks.
The reasoning is then very similar  to that for the Drell-Yan process
discussed above.  Note in particular,
that the partonic subreaction remains an annihilation (or gluon fusion)
process.  

To apply our methods to heavy
quark and other QCD production mechanisms,
Eq.\ (\ref{qqgammastarfact}) must be generalized in 
a number of ways.  First, the underlying
hard process involves not only a singlet color configuration, but
any color tensor that may be constructed from the 
color representations of the incoming partons.  Second,
the outgoing heavy quarks (or other particles), as well as the
incoming light partons, act as sources of soft gluons.
Our analysis is simplified, however, by the following
observation.  Near partonic threshold, the emission of
light quarks into the final state is suppressed by
factors of $1-z$, aside from pairs produced by 
soft gluons.  As a result, although different color
structures may ``mix" due to soft gluon emission, the
{\it flavor} structure of the hard scattering is 
unchanged by all corrections that are singular in $1-z$.
 Thus in the case
of heavy quark production, we may analyze light
quark annihilation and gluon fusion separately.

To take these new features into account, we generalize 
the  Drell-Yan factorization to 
QCD hard scattering initiated by
light partons $f$ and $\bar f$
with the following expressions, which are the analogs of 
Eqs.\ 
(\ref{dymoment}) (for $h_1=f$, $h_2={\bar f})$ and 
(\ref{qqgammastarfact}),
\beqa
\int_0^1 d\tau \tau^{N-1}
{d\sigma_{f{\bar f}\rightarrow Q{\bar Q}} \over d Q^2}(\tau, Q^2,\epsilon)
&\ & 
\cr
&\ & \hskip -1.5 true in
=\ {\tilde \phi}_{f/f}(N,\mu^2,\epsilon)\; 
{\hat \sigma}_{f{\bar f}\rightarrow Q{\bar Q}}(N,Q/\mu,Q,\alpha_s(\mu^2))\; 
{\tilde \phi}_{{\bar f}/{\bar f}}(N,\mu^2,\epsilon)
\cr
&\ & \hskip -1.5 true in 
= \sum_{IJ} 
h^*_J\left({Q\over\mu},\alpha_s(\mu^2)\right)\; 
{\tilde S}_{JI} \biggl ( {Q\over \mu N},\alpha_s(\mu^2) \biggr )\; 
h_I\left({Q\over\mu},\alpha_s(\mu^2)\right) \cr
&\ & \hskip -1.25 true in\times
{\tilde\psi}_{f/f}\left ( N,{Q\over \mu },\epsilon \right )
\;
{\tilde\psi}_{{\bar f}/{\bar f}}\left (N,{Q\over \mu },\epsilon \right )  
+{\cal O}(1/N)\, .
\label{qqQQfact}
\eeqa
This factorization is illustrated in Fig.\ 2, for
the process
\beq
f(p_a)+{\bar f}(p_b)\rightarrow {\bar Q}(p_1) + Q(p_2)\, ,
\label{ffbartoQQbar}
\eeq
in which $f{\bar f}$ represents a pair of light quarks
or gluons that annihilate or fuse, respectively, into a heavy quark pair.

\begin{figure}
\centerline{
\psfig{file=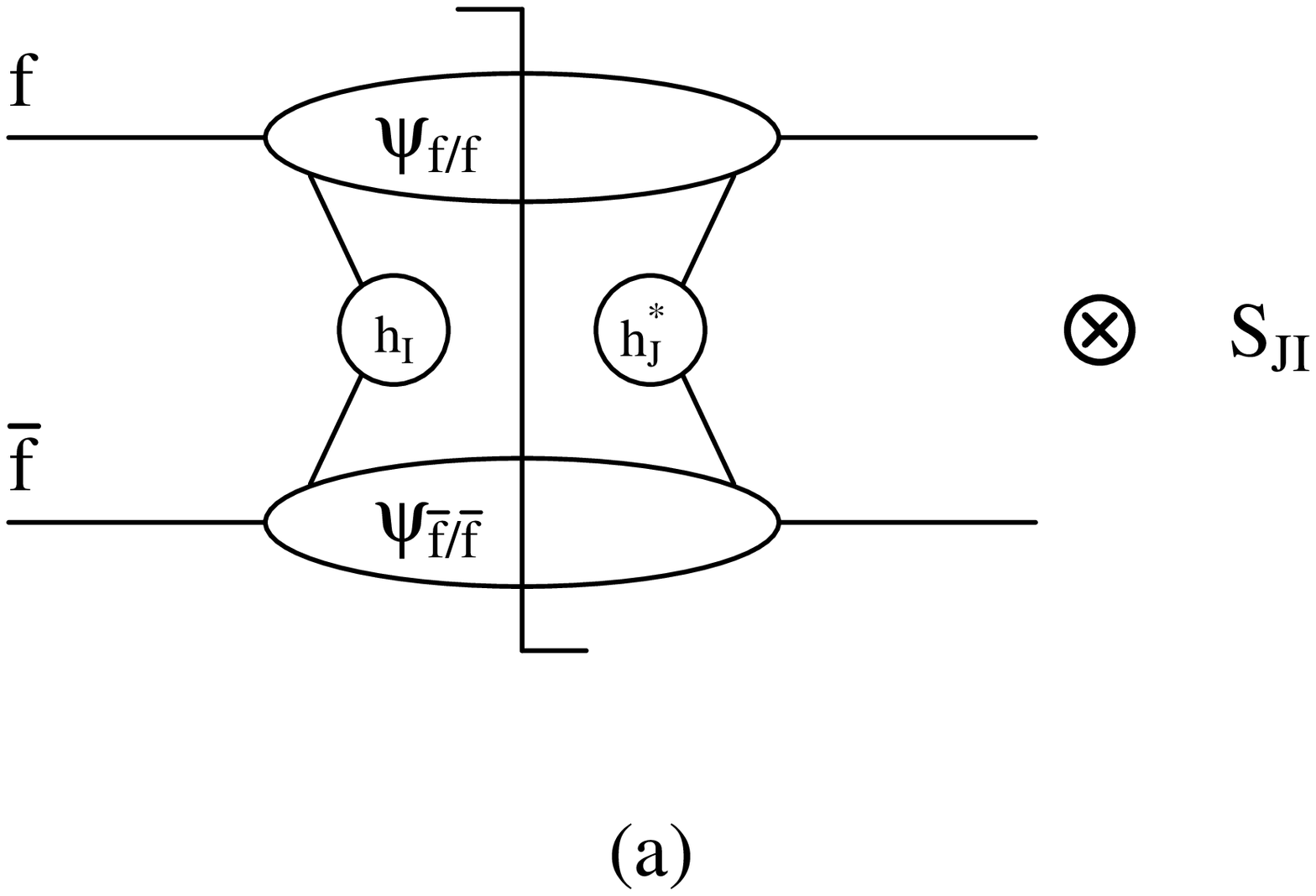,height=2.5in,width=4.05in,clip=}}
\vspace{5mm}
\centerline{
\psfig{file=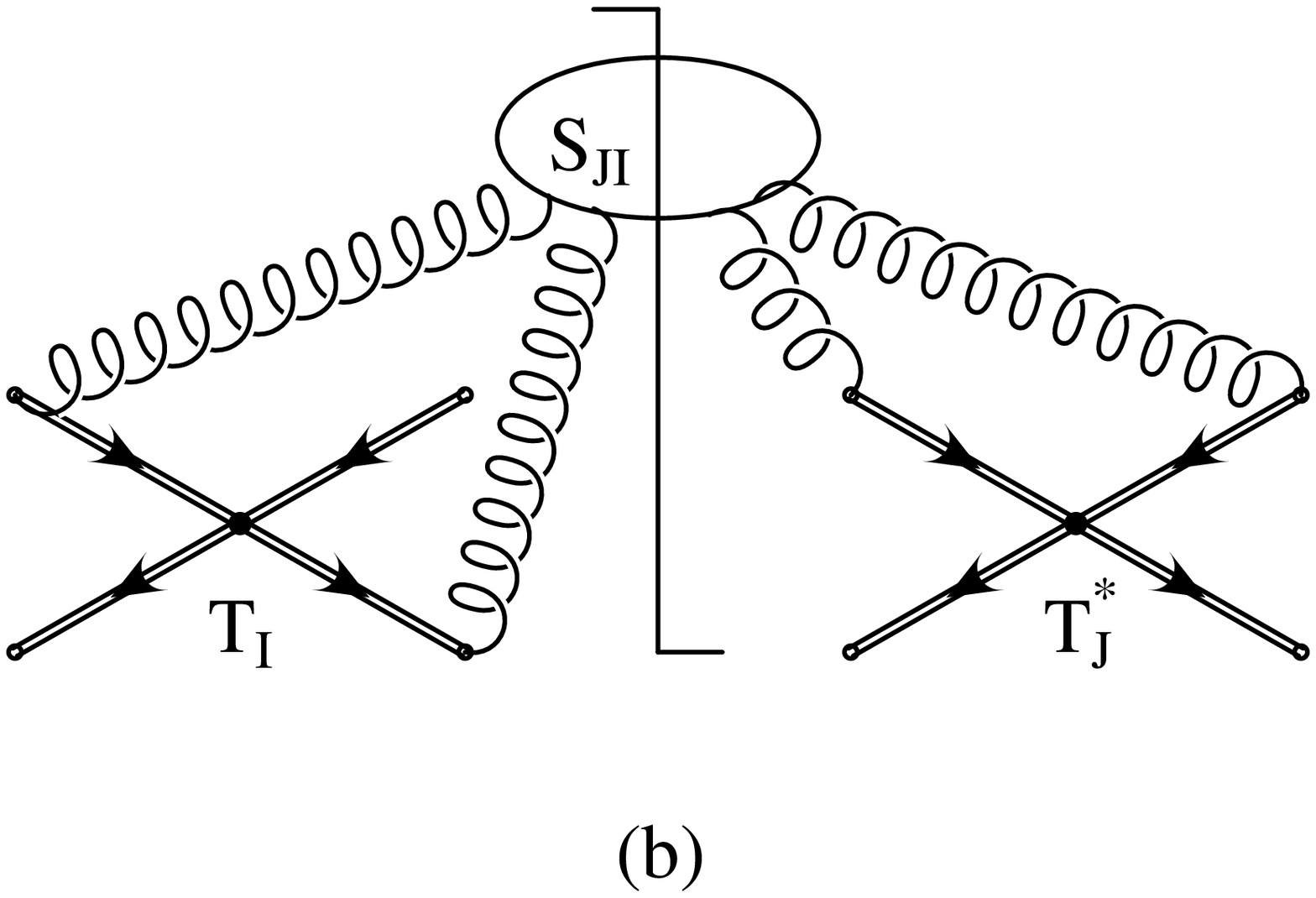,height=2.5in,width=4.05in,clip=}}
{Fig. 2. (a) Factorization for heavy quark production near partonic threshold,
Eq.\ (\ref{qqQQfact}). (b) The soft-gluon function $S_{JI}$, in which
the vertices $T$ link ordered exponentials as in Eq.\ (\ref{wIdef}).}
\label{fig. 2}
\end{figure}

This factorization differs from Eq.\ (\ref{qqgammastarfact}) only
in taking into account color exchange at the hard 
scatterings, $h_I$ in the amplitude, and $h^*_J$ in
the complex conjugate.  For simplicity, we omit the
partonic indices $f,{\bar f}$ in $h_I$ and $h^*_J$,
although these functions certainly depend on incoming flavor.
For quark-antiquark annihilation, the indices $I$ and $J$ range over
two values, corresponding to annihilation in a singlet or
octet in the $s$-channel (or $t$ channel). 
For other hard-scatterings, such as gluon-gluon, the same
pattern applies, but with $I$ and $J$ ranging over
color tensors appropriate to possible combinations
of gluonic representations in the initial state with
(heavy) quark representations in the final state.

The color tensor of the hard scattering serves to
tie together the ordered exponentials to which soft
gluons couple, generalizing the singlet vertex joining
the eikonal lines in Eq.\ (\ref{WDY}) above,
\beqa
w_I(x)_{\{c_k\}}
&=& 
\Phi_{v_j}(\infty,0;x)_{c_j,d_j}\; 
\Phi_{v_i}(0,\infty;x)_{d_i,c_i}\cr
&\ &\times
\left( T_I\right)_{d_bd_a,d_jd_i}\; 
\Phi_{v_a}(-\infty,0;x)_{c_a,d_a}
\Phi_{v_b}(0,-\infty;x)_{d_b,c_b}\, ,\cr
&\ &
\label{wIdef}
\eeqa
where $\Phi_{v_a}$ is in the color representation
flavor $f$, with velocity $v_a$.
Notice that only the Wilson lines of the incoming
partons have light-like velocity vectors.
  Next, by analogy to Eq.\ (\ref{udef})
for the eikonal cross section in Drell-Yan, we define,
\beqa
\sigma^{({\rm eik})}_{JI}\left((1-w)Q/\mu,\alpha_s(\mu^2)\right)
&=&{Q\over 2\pi}\;
\int d x_0\; {\rm e}^{i(1-w)Qx_0}\; 
\langle 0|{\bar T}\left[ w^\dagger_J(x_0,\vec{0})_{\{b_i\}}\right]\cr
&\ & \quad \quad \times
T\left[w_I(0)_{\{b_i\}}\right]|0\rangle\, .
\label{sigijdef}
\eeqa
As in the Drell-Yan case, this cross section includes
unphysical collinear divergences associated with its
incoming, lightlike eikonal lines.  These are eliminated
in the soft function $S_{JI}$ by refactorizing
$\sigma^{\rm (eik)}_{JI}$ in the same manner 
as $\sigma^{\rm (eik)}_{\rm DY}$, Eq.\ (\ref{sigeikDY}),
through moments,
\beq
{\tilde \sigma}^{({\rm eik})}_{JI}(N)
=
{\tilde S}_{JI}(N)\, {\tilde j}_1(N)\; {\tilde j}_2(N)\, .
\label{sigeik}
\eeq

Now, comparing Eqs.\ (\ref{qqgammastarfact}) and (\ref{qqQQfact}), we 
see that the $N$-dependence of the cross section
for heavy-quark production from a set of initial-state
partons is given by the following generalization of Eq.\ (\ref{omegaqqofN}),
\begin{equation}
{\hat \sigma}_{f{\bar f}\rightarrow Q{\bar Q}}(N)
=
\left[{{\tilde\psi}_{f/f}(N,1,\epsilon)\over{\tilde \phi}_{f/f}(N,Q^2,\epsilon)}\right]^2\,  
\sum_{IJ} h^*_J(Q / \mu)\; {\tilde S}_{JI}(Q / N\mu)\; h_I(Q /\mu)\, ,
\label{hatsigtoqqbar}
\end{equation}
where the sum over $I$ and $J$ is, as usual, over color tensors.
To determine the singular behavior at partonic threshold, 
we only need to 
resum logarithmic $N$-dependence in each $S_{JI}$.  

The soft matrix $S_{JI}$ depends on $N$ through the ratio $Q/(N\mu)$.
Its $N$-dependence, then, can be resummed by renormalization group
analysis \cite{bbj,BottsSt,GK,KK,Lipatov}.
The composite operator in Eq.\ (\ref{wIdef}), which defines $S_{JI}$,
requires renormalization.
Nevertheless, the product $h_J^*S_{JI}h_I$
of the soft function and the hard factors needs
no overall renormalization, because the UV divergences of $S_{JI}$
associated with the new composite operator are balanced by those
of the $h$'s.  This situation is summarized by the conditions
\cite{BottsSt,CLS} 
\beqa
h^{(0)}_I&=& \prod_{i=1}^2Z_i^{-1/2}\; \left(Z_S^{-1}\right)_{IC}h_{C}\cr
{h^*}^{(0)}_J&=& \prod_{i=1}^2Z_i^{-1/2}\; h^*_D\; [(Z_S^\dagger)^{-1}]_{DJ}\cr
S^{(0)}_{JI}&=&(Z_S^\dagger)_{JB}S_{BA}Z_{S,AI},
\label{hSrenorm}
\eeqa
where $h^{(0)}$ and $S^{(0)}$ denote the unrenormalized quantities,
$Z_i$ is the renormalization constant of the $i$th
incoming partonic field external to $h_I$ and $Z_{S,IJ}$ is
a matrix of renormalization constants, which describe the
renormalization of the soft function, including
mixing of color structures.  $Z_{S}$ is defined to include the
wave function renormalization 
necessary for the outgoing eikonal lines that
represent the heavy quarks.

From Eq.\ (\ref{hSrenorm}), the soft function  $S_{JI}$ satisfies the
renormalization group equation \cite{BottsSt}
\begin{equation}
\left(\mu {\partial \over \partial \mu}+\beta(g){\partial \over \partial g}
\right)\,S_{JI}
=-(\Gamma^\dagger_S)_{JB}S_{BI}-S_{JA}(\Gamma_S)_{AI}\, ,
\label{gammaRG}
\end{equation}
where the anomalous dimension matrices may be
found by explicit renormalization of the soft function.
Following \cite{BottsSt}, we  find it convenient to compute
the matrix
in a minimal subtraction scheme.
To be explicit, taking $\epsilon=4-n$ the matrix of
anomalous dimensions may be found 
at one loop from the matrix of 
renormalization constants by
\begin{equation}
\Gamma_S (g)=-\frac{g}{2} \frac {\partial}{\partial g}{\rm Res}_{\epsilon 
\rightarrow 0} Z_S (g, \epsilon).
\label{residuecalc}
\end{equation}
The determination of the anomalous dimension
matrices $\Gamma_S$ will be the subject of the following
sections, for heavy quark production
through the annihilation of light quarks and  gluon fusion.

The solution to the renormalization group equation (\ref{gammaRG})
organizes logarithms of $N$ in the soft function.  In general, this
matrix equation involves a scale-dependent mixing of tensors, and
has no closed expression.  At the level of leading logarithms of $N$
in $S_{JI}$, and therefore at next-to-leading logarithm of $N$ in
the cross section as a whole, we may simplify by
choosing a color basis in which the anomalous dimension matrix is
diagonal, with eigenvalues $\lambda_I$ for each basis color
tensor labelled by $I$.
At this level of approximation, we have
\begin{eqnarray}
{\tilde S}_{JI}\left(\frac{Q}{N\mu_2}, \, \alpha_s(\mu_2^2)\right)&=&
{\tilde S}_{JI}\left(\frac{Q}{N\mu_1}, \, \alpha_s\left(\mu_1^2\right)\right)
\nonumber \\ 
&\ & \quad \times 
\exp\left[-\int^{\mu_2}_{\mu_1}\frac{d \bar{\mu}}{\bar{\mu}}
[\lambda_I(\alpha_s(\bar{\mu}^2))+\lambda^*_J(\alpha_s(\bar{\mu}^2))]\right]\, .
\cr &\ &
\label{SIJRGsoln}
\end{eqnarray}
The coherent effects of soft gluon emission are all contained
in this expression, which we now employ to determine ${\hat \sigma}_{f{\bar f}\rightarrow Q{\bar Q}}$,
Eq.\ (\ref{hatsigtoqqbar}), to NLL accuracy.

Our first step is to rewrite  Eq.\ (\ref{hatsigtoqqbar}) slightly, to
isolate differences and similarities to the Drell-Yan result
of Eq.\ (\ref{omegaqqofN}),
\beqa
{\hat \sigma}_{f{\bar f}\rightarrow Q{\bar Q}}(N)
&=&
\left[{{\tilde \psi}_{f/f}(N,1,\epsilon)\over{\tilde \phi}_{f/f}(N,Q^2,\epsilon)}\right]^2\;
{\tilde U}^{(f)}(Q/N\mu)\cr  
&\ & \quad \times
{\sum_{IJ} h^*_J(Q / \mu)\; {\tilde S}_{JI}(Q / N\mu)\; h_I(Q /\mu)\over {\tilde U}^{(f)}(Q/N\mu)}\, .
\label{hatsigmodified}
\eeqa
We have simply multiplied and divided by $U^{(f)}(Q/N\mu)$,
defined in Eq.\ (\ref{sigeikDY}).   
The first factor in Eq.\ (\ref{hatsigmodified}) is just the
resummed singlet cross section (\ref{omegaexp}).  The new factor
is to be treated using Eq.\ (\ref{SIJRGsoln}) and the explicit
expression of $U$, Eq.\ (\ref{Uresum}).

Combining the results just discussed, and picking a color basis 
in which moments with respect to $z$ exponentiate, we find, at NLL accuracy
\begin{eqnarray}
{\hat \sigma}_{f{\bar f}\rightarrow Q{\bar Q}}(N) &\ &\nonumber \\ 
&\ & \hskip - 1.0 true in
= A'(\alpha_s(Q^2))\; h^*_J\left(1,\alpha_s(Q^2)\right)\; 
{\tilde S}_{JI} \biggl ( 1,\alpha_s([Q/N]^2) \biggr )\; 
h_I\left(1,\alpha_s(Q^2)\right) 
\nonumber \\
&\ & \hskip - 1.0 true in
\quad \quad \times
\exp \left[ E_{JI}(N,\theta,Q^2)\right ] \, ,
\label{omegaofn}
\end{eqnarray}
where $A'$ is an overall constant and
where the color-dependent exponents are given by
\begin{eqnarray}
E_{JI}(N,\theta,Q^2)&=&-\int_0^1 dz \frac{z^{N-1}-1}{1-z}
 \biggl [\int^{(1-z)^{m_S}}_{(1-z)^2} \frac{d\lambda}{\lambda} 
g_1^{(ab)}[\alpha_s(\lambda Q^2)]
\nonumber \\ &&
+\, g_2^{(ab)}[\alpha_s((1-z)^{m_S} Q^2)]
\nonumber \\ &&
+g_3^{(I)}[\alpha_s((1-z)^2 Q^2),\theta] 
+g_3^{(J)*}[\alpha_s((1-z)^2 Q^2),\theta]\, \biggr ]\, .
\nonumber \\
\label{Eofn}
\end{eqnarray}
$g_3^{(I)}$ summarizes
soft logarithms that depend directly on color exchange in the hard scattering,
and hence also on the identities and relative directions of the colliding 
partons (through $\theta$), both incoming and outgoing.
As pointed out above, one-loop contributions to
$g_3$ may always be absorbed into the one-loop contribution to $g_2$
and the two-loop contribution to $g_1$ \cite{CT2}. Because
$g_3^{(I)}$ depends upon $I$, however, it is advantageous to keep
this nonfactoring color-dependence separate.

Given a choice of incoming and outgoing partons, 
next-to-leading logarithms
in the moment variable $N$ exponentiate as in (\ref{omegaofn}) in the color 
tensor
basis that diagonalizes $\Gamma^{(ab\rightarrow cd)}_{S, JI}$,
with eigenvalues $\lambda_I$.  The
resulting soft gluon anomalous dimension $g_3^{(I)}$ is then simply
\begin{equation}
g_3^{(I)}[\alpha_s,\theta]=-\lambda_I[\alpha_s,\theta]
+{1\over 2}\nu^{(i)}(\alpha_s)\, ,
\label{gthreedef}
\end{equation}
where the eigenvalues are complex in general, and depend
on the relative directions of the incoming and outgoing partons, as shown.

\mysection{Applications to $q \bar{q} \rightarrow Q \bar{Q}$}

These considerations may be illustrated by heavy quark
production through light quark annihilation,
\begin{equation}
q(p_a)+{\bar q}(p_b) \rightarrow {\bar Q}(p_1) + Q(p_2)\, .
\end{equation}
In this case,
as in quark-quark elastic scattering \cite{BottsSt,KK}, the anomalous
dimension matrix is two-dimensional.  
Following Ref.\ \cite{mengetal} we define the invariants 
\begin{equation}
s=(p_a+p_b)^2\, , \quad t_1=(p_a-p_1)^2-m^2\, , \quad u_1=(p_b-p_1)^2-m^2\, , 
\end{equation}
with $m$ the heavy quark mass, which satisfy
\begin{equation}
s+t_1+u_1=0
\end{equation}
at partonic threshold. 
We also define dimensionless vectors $v_i^{\mu}$ by
\begin{equation}
p_i^{\mu}={\cal Q}v_i^{\mu}
\end{equation}
which obey $v_i^2=0$ for the light incoming quarks and
$v_i^2=m^2/{\cal Q}^2$ for the outgoing heavy quarks.
Note that $\cal Q$ satisfies the kinematic relation
$s=2{\cal Q}^2$.

We now calculate the matrix $\Gamma_S (g)$ for this case. 
Renormalization associated with incoming light partons
is factorized into the wave functions $\psi$ of the
previous section.  We therefore need consider only
diagrams which are not part of the eikonal ``jets"
$j_i$, as in Eq.\ (\ref{eikjdef}).  In particular, we
may omit the UV structure of the incoming light-like
eikonal lines.

The UV divergent $O(\alpha_s)$ contribution to 
each $S_{JI}$ is the sum of  
graphs illustrated by Fig.\ 3.   Most interesting
are the ``vertex corrections" in Fig. 3(a).
The counterterms for $S$ are the ultraviolet divergent
coefficients times our basis color tensors, denoted $c_i$
\begin{eqnarray}
S_1&=&c_1 Z_{S, 11} + c_2 Z_{S, 21}, \nonumber
\\
S_2&=&c_1 Z_{S, 12} + c_2 Z_{S, 22}\, ,
\end{eqnarray}
where we will specify a convenient basis for the $c_i$ below.

\begin{figure}
\centerline{
\psfig{file=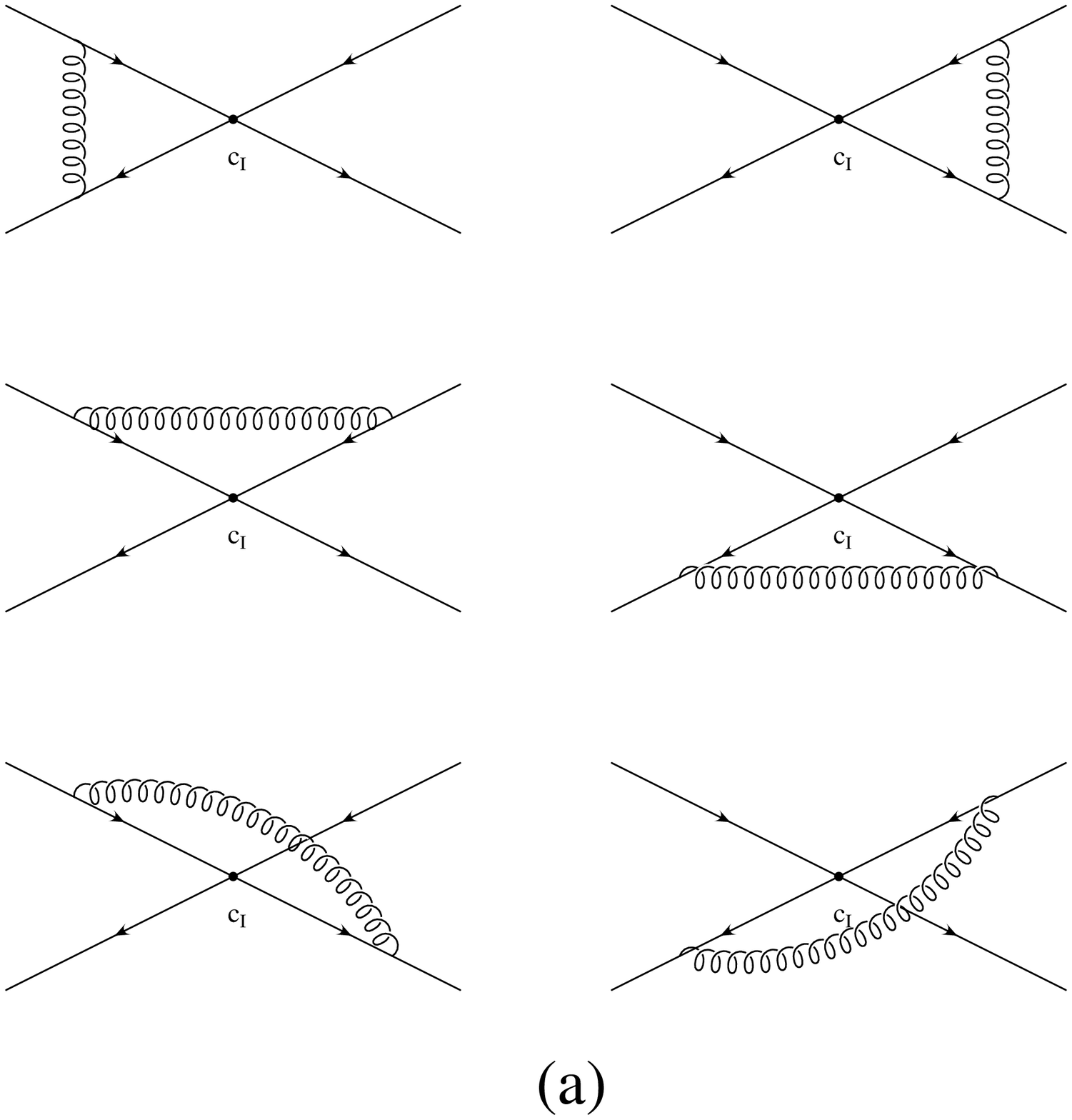,height=4.05in,width=4.05in,clip=}}
\vspace{13mm}
\centerline{
\psfig{file=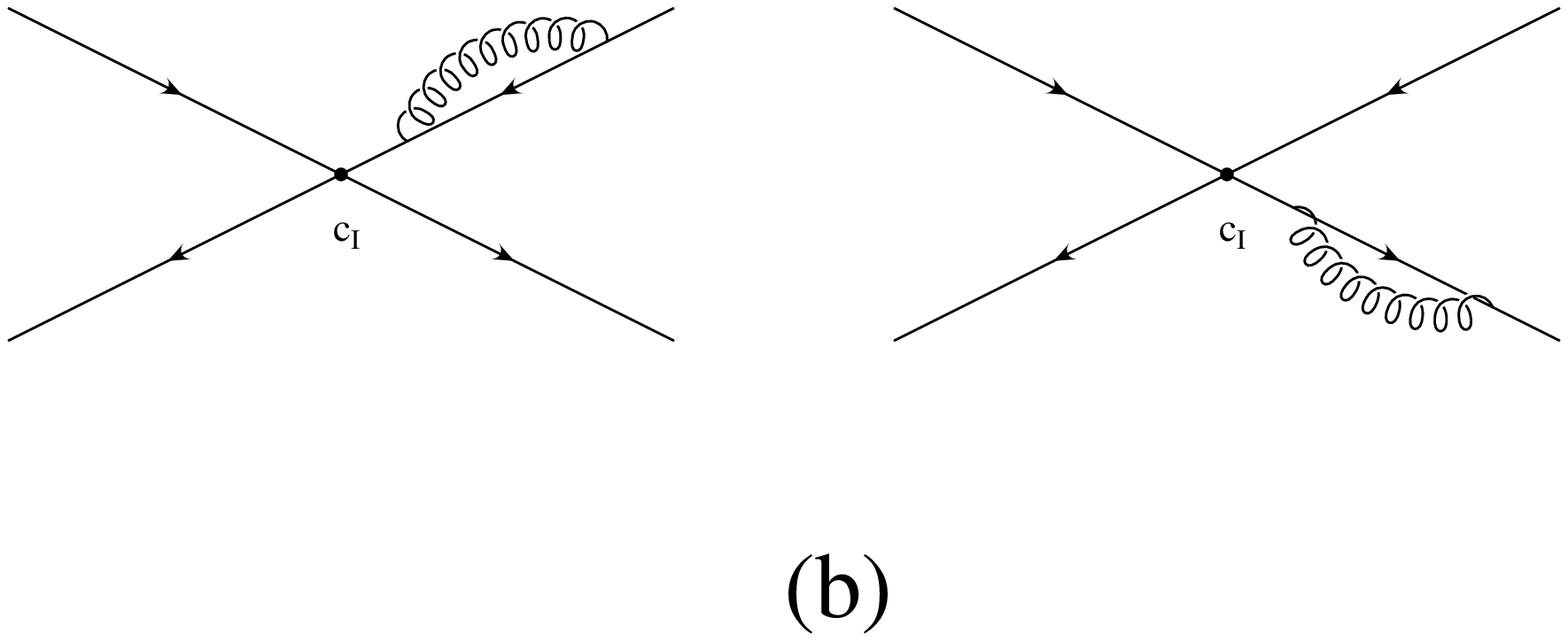,height=1.5in,width=4.05in,clip=}}
{Fig. 3. One-loop corrections to $S_{JI}$ for heavy quark production
through light quark annihilation: (a) vertex corrections;
(b) heavy quark self-energy graphs.}
\label{fig. 3}
\end{figure}

In our calculations we 
employ the general axial gauge gluon propagator,
\begin{equation}
D^{\mu \nu}(k)=\frac{-i}{k^2+i\epsilon} N^{\mu \nu}(k), \quad
N^{\mu \nu}(k)=g^{\mu \nu}-\frac{n^{\mu}k^{\nu}+k^{\mu}n^{\nu}}{n \cdot k}
+n^2\frac{k^{\mu}k^{\nu}}{(n \cdot k)^2},
\end{equation}
with $n^{\mu}$ the gauge vector,
and eikonal rules for all external lines (Fig.\ 4).
We now review the notation of  Ref.\ \cite{BottsSt},
which enables us to summarize results for the
diagrams of Fig.\ 3 in a unified manner.

\begin{figure}
\centerline{\hspace{13mm}
\psfig{file=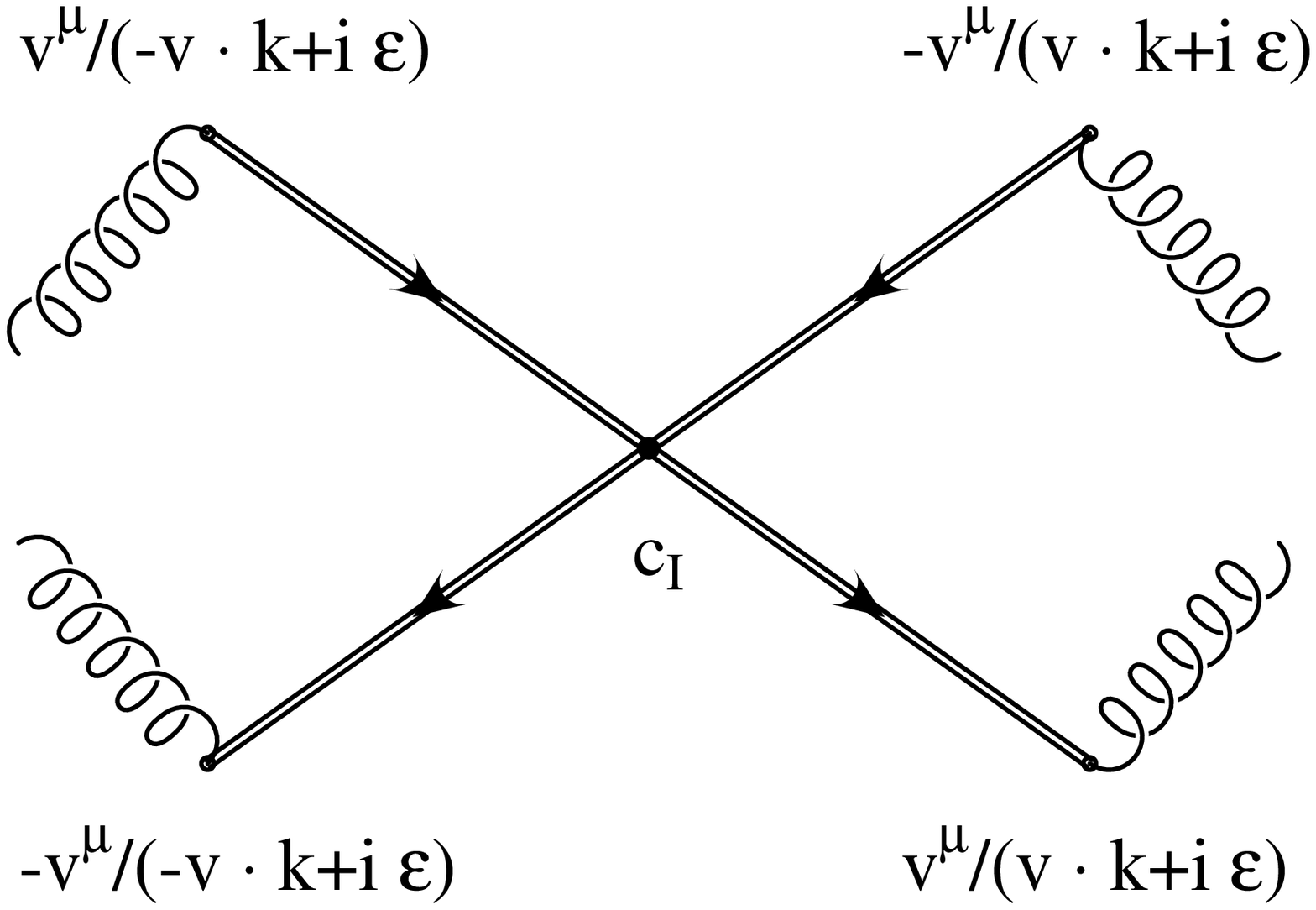,height=2.5in,width=4.05in,clip=}}
{Fig. 4. Feynman rules for eikonal lines 
in $q {\bar q} \rightarrow Q {\bar Q}$. 
The gluon momentum flows out of the eikonal lines.
Group matrices at the vertices are the same as
for quark lines.}
\label{fig. 4}
\end{figure}

If we denote a typical one-loop correction to $c_I$ as
$\omega^{(I)}(\delta_i v_i, \delta_j v_j, \Delta_i, \Delta_j)$,
where $\delta=+1 \,(-1)$ for the gluon momenta flowing in the same (opposite)
direction to the momentum of $v_i$ and $\Delta=+1 \, (-1)$ for $v_i$ 
corresponding to a quark (antiquark),
then we have:
\begin{eqnarray}
\omega^{(I)}&=& c_I \, g^2
\int\frac{d^n q}{(2\pi)^n}\frac{-i}{q^2+i\epsilon}
\left\{\frac{\Delta_i\Delta_j v_i \cdot v_j}{(\delta_i v_i\cdot q+i\epsilon)
(\delta_j v_j\cdot q+i\epsilon)}\right.
\nonumber \\ &&
\left. -\frac{\Delta_i v_i \cdot n}{(\delta_i v_i\cdot q+i\epsilon)}
\frac{P}{(n \cdot q)} 
-\frac{\Delta_j v_j \cdot n}{(\delta_j v_j\cdot q+i\epsilon)}
\frac{P}{(n \cdot q)}
+n^2\frac{P}{(n \cdot q)^2}\right\},
\nonumber \\
\label{omega}
\end{eqnarray}
where $c_I$ is a color tensor.
$P$ stands for principal value,
\begin{equation}
\frac{P}{(q \cdot n)^{\beta}}=\frac{1}{2}\left(\frac{1}{(q \cdot n+i\epsilon)
^{\beta}}+(-1)^{\beta}\frac{1}{(-q \cdot n+i\epsilon)^{\beta}}\right).
\end{equation}
We rewrite (\ref{omega}) as
\begin{eqnarray}
\omega^{(I)}&=&c_I \, {\cal S}_{ij} 
\left[I_1(\delta_i v_i, \delta_j v_j)
-\frac{1}{2}I_2(\delta_i v_i, n)-\frac{1}{2}I_2(\delta_i v_i, -n)\right.
\nonumber \\ && \quad \quad
\left.-\frac{1}{2}I_3(\delta_j v_j, n)-\frac{1}{2}I_3(\delta_j v_j, -n)
+I_4(n^2)\right] \, ,
\label{Isum}
\end{eqnarray}
where the overall sign is given by
\begin{equation}
{\cal S}_{ij}=\Delta_i \: \Delta_j \: \delta_i \: \delta_j.
\end{equation}

We now evaluate the ultraviolet poles of the integrals.
For the integrals when both $v_i$ and $v_j$ refer to massive quarks we have
(with $\epsilon=4-n$)
\begin{eqnarray}
I_1^{UV \;pole}&=&\frac{\alpha_s}{\pi}\frac{1}{\epsilon} L_{\beta} \, , 
\nonumber\\ 
I_2^{UV \;pole}&=&-\frac{\alpha_s}{\pi}\frac{1}{\epsilon} L_i \, , 
\nonumber\\
I_3^{UV \;pole}&=&-\frac{\alpha_s}{\pi}\frac{1}{\epsilon} L_j \, , 
\nonumber\\
I_4^{UV \;pole}&=&-\frac{\alpha_s}{\pi}\frac{1}{\epsilon} \, ,
\end{eqnarray}
where the $L_\beta$ is the familiar velocity-dependent
eikonal function
\begin{equation}
L_{\beta}=\frac{1-2m^2/s}{\beta}\left(\ln\frac{1-\beta}{1+\beta}
+\pi i \right)\, ,
\end{equation}
with $\beta=\sqrt{1-4m^2/s}$.
The $L_i$ and $L_j$ are rather complicated functions of
the gauge vector $n$. We will see shortly that their
contributions are cancelled by the inclusion of self energies. 
Their explicit expressions are:
\begin{equation}
L_i=\frac{1}{2}[L_i(+n)+L_i(-n)] \, ,
\label{Ellidef}
\end{equation}
where
\begin{eqnarray}
L_i(\pm n)&=&\frac{1}{2}\frac{|v_i \cdot n|}{\sqrt{(v_i \cdot n)^2-2m^2n^2/s}}
\nonumber \\ &&
\left[\ln\left(\frac{\delta(\pm n) \, 2m^2/s-|v_i \cdot n| 
- \sqrt{(v_i \cdot n)^2-2m^2n^2/s}}
{\delta(\pm n) \, 2m^2/s-|v_i \cdot n| 
+ \sqrt{(v_i \cdot n)^2-2m^2n^2/s}}\right)\right.
\nonumber \\ && 
+\left.\ln\left(\frac{\delta(\pm n) \, n^2-|v_i \cdot n| 
- \sqrt{(v_i \cdot n)^2-2m^2n^2/s}}
{\delta(\pm n) \, n^2-|v_i \cdot n| 
+ \sqrt{(v_i \cdot n)^2-2m^2n^2/s}}\right)\right] 
\end{eqnarray}
with $\delta(n) \equiv |v_i \cdot n|/ (v_i \cdot n)$.

When $v_i$ refers to a massive quark and $v_j$ to a massless quark we have
\begin{eqnarray}
I_1^{UV \;pole}&=&\frac{\alpha_s}{2\pi}
\left\{\frac{2}{\epsilon^2}-\frac{1}{\epsilon}
\left[\gamma+\ln\left(\frac{v_{ij}^2 s}{2m^2}\right)-\ln(4\pi) \right]
\right\}\, , \nonumber\\ 
I_2^{UV \;pole}&=&-\frac{\alpha_s}{\pi}\frac{1}{\epsilon} L_i \, , 
\nonumber\\
I_3^{UV \;pole}&=&\frac{\alpha_s}{2\pi}
\left\{\frac{2}{\epsilon^2}-\frac{1}{\epsilon}
\left[\gamma+\ln(\nu_j)-\ln(4\pi)\right]
\right\} \, , 
\nonumber\\
I_4^{UV \;pole}&=&-\frac{\alpha_s}{\pi}\frac{1}{\epsilon} \, , 
\end{eqnarray}
where
\begin{equation}
\nu_a=\frac{(v_a \cdot n)^2}{|n|^2}\, ,
\end{equation}
and $v_{ij}=v_i \cdot v_j$. Note that the double poles cancel
in the sum over the $I$'s in Eq.\ (\ref{Isum}).

Finally, when both $v_i$ and $v_j$ refer to massless quarks we have 
\cite{BottsSt}
\begin{eqnarray}
I_1^{UV \;pole}&=&\frac{\alpha_s}{\pi}
\left\{\frac{2}{\epsilon^2}-\frac{1}{\epsilon}
\left[\gamma+\ln\left(\delta_i\delta_j\; \frac{v_{ij}}{2}\right)
-\ln(4\pi) \right]\right\} \, , 
\nonumber\\ 
I_2^{UV \;pole}&=&\frac{\alpha_s}{2\pi}
\left\{\frac{2}{\epsilon^2}-\frac{1}{\epsilon}
\left[\gamma+\ln(\nu_i)-\ln(4\pi)\right]\right\} \, , 
\nonumber\\
I_3^{UV \;pole}&=&\frac{\alpha_s}{2\pi}
\left\{\frac{2}{\epsilon^2}-\frac{1}{\epsilon}
\left[\gamma+\ln(\nu_j)-\ln(4\pi)\right]\right\} \, , 
\nonumber\\
I_4^{UV \;pole}&=&-\frac{\alpha_s}{\pi}\frac{1}{\epsilon}\, . 
\end{eqnarray}
Again, note that the double poles cancel.

Our calculations 
are most easily carried out in a color tensor basis
consisting of singlet exchange in the $s$ and $u$ channels,
\begin{equation}
c_1 = \delta_{ab}\delta_{12}\, , \quad \quad
c_2 =  \delta_{a2}\delta_{b1}\, .
\label{alter}
\end{equation}
The color indices for the incoming (light) quark and antiquark are $a$ and $b$,
respectively, and for the outgoing (heavy) quark and antiquark $2$ and $1$, 
respectively.   

In the basis (\ref{alter}) we find from Eq.\ (\ref{residuecalc}) that
the contributions of the diagrams in Fig. 3(a) to the anomalous dimensions are
\begin{eqnarray}
{\hat\Gamma}_{S, 11}&=&\frac{\alpha_s}{\pi}C_F\left[\ln
\left(\frac{v_{ab}}{2}\right)-L_{\beta}
-\frac{1}{2}\ln(\nu_a\nu_b)-L_1-L_2+2-\pi i\right]
\nonumber \\ &&
-\frac{1}{N}{\hat\Gamma}_{S, 21},
\nonumber \\
{\hat\Gamma}_{S, 21}&=&\frac{\alpha_s}{2\pi}
\ln\left(\frac{v_{a2}v_{b1}}{v_{a1}v_{b2}}\right),
\nonumber \\ 
{\hat\Gamma}_{S, 12}&=&\frac{\alpha_s}{2\pi}
\left[\ln\left(\frac{v_{ab}}{2}\right)-\ln(v_{a1}v_{b2})-L_{\beta}
+\ln\left(\frac{2m^2}{s}\right)-\pi i\right],
\nonumber \\
{\hat\Gamma}_{S, 22}&=&\frac{\alpha_s}{\pi}C_F
\left[\ln(v_{a2}v_{b1})-\ln\left(\frac{2m^2}{s}\right)
-\frac{1}{2}\ln(\nu_a\nu_b)-L_1-L_2+2\right]
\nonumber \\ &&
-\frac{1}{N}{\hat\Gamma}_{S, 12}.
\label{hatGamma1loop}
\end{eqnarray}
The matrix depends, as expected, on the directions of the Wilson
lines, which may be reexpressed in terms of ratios of
kinematic invariants for the partonic scattering.  It is also
gauge-dependent.

We now combine this matrix with the (color-diagonal)
 contributions from
the heavy-quark
self-energy graphs in Fig.\ 3(b).
The contribution of the self-energy graphs is 
\begin{equation}
\frac{\alpha_s}{\pi} C_F(L_1+L_2-2)\; \delta_{IJ}\, ,
\end{equation}
which cancels the gauge-dependent $L_i$ of Eq.\ (\ref{Ellidef}).
In addition, recalling Eq.\ (\ref{gthreedef}),
 we subtract half of the anomalous dimension 
 $\nu^{(i)}=2(\alpha_s/\pi)C_{F,A}$ 
associated with the soft gluon function $U^{(i)}$ of the 
Drell-Yan cross section 
(see Eq.\ (\ref{hatsigmodified})).
This shifts our matrix by an additional $-(\alpha_s/\pi)C_F\delta_{IJ}$. 
Combining these contributions with 
${\hat \Gamma}_S$ of Eq. (\ref{hatGamma1loop}), 
the full soft-gluon anomalous dimension matrix becomes
\begin{eqnarray}
\Gamma_{S, 11}&=&\frac{\alpha_s}{\pi}C_F\left[\ln
\left(\frac{v_{ab}}{2}\right)-L_{\beta}-\frac{1}{2}\ln(\nu_a\nu_b)-1
-\pi i\right]
\nonumber \\ &&
-\frac{1}{N}\Gamma_{S, 21},
\nonumber \\
\Gamma_{S, 21}&=&\frac{\alpha_s}{2\pi}
\ln\left(\frac{v_{a2}v_{b1}}{v_{a1}v_{b2}}\right),
\nonumber \\
\Gamma_{S, 12}&=&\frac{\alpha_s}{2\pi}
\left[\ln\left(\frac{v_{ab}}{2}\right)-\ln(v_{a1}v_{b2})-L_{\beta}
+\ln\left(\frac{2m^2}{s}\right)-\pi i\right],
\nonumber \\
\Gamma_{S, 22}&=&\frac{\alpha_s}{\pi}C_F
\left[\ln(v_{a2}v_{b1})-\ln\left(\frac{2m^2}{s}\right)
-\frac{1}{2}\ln(\nu_a\nu_b)-1\right]
\nonumber \\ &&
-\frac{1}{N}\Gamma_{S, 12}\, .
\end{eqnarray}

In the $A^0=0$ gauge we have $\nu_a=\nu_b=1/2$  in the partonic
center of mass frame.
Using the relations $v_{ab}=1$, $v_{12}=1-2m^2/s$, $v_{a1}=v_{b2}=-t_1/s$
and $v_{1b}=v_{2a}=-u_1/s$, the anomalous dimension matrix
becomes in terms of our invariants $s$, $t_1$, and $u_1$ :
\begin{eqnarray}
\Gamma_{S, 11}&=&\frac{\alpha_s}{\pi}\left\{C_F\left[\ln
\left(\frac{u_1^2}{t_1^2}\right)-L_{\beta}-1-\pi i\right]-\frac{C_A}{2}
\ln\left(\frac{u_1^2}{t_1^2}\right)\right\},
\nonumber \\
\Gamma_{S, 21}&=&\frac{\alpha_s}{2\pi}
\ln\left(\frac{u_1^2}{t_1^2}\right),
\nonumber \\ 
\Gamma_{S, 12}&=&\frac{\alpha_s}{2\pi}
\left[\ln\left(\frac{m^2 s}{t_1^2}\right)-L_{\beta}-\pi i\right],
\nonumber \\
\Gamma_{S, 22}&=&\frac{\alpha_s}{\pi}\left\{C_F
\left[\ln\left(\frac{u_1^2}{t_1^2}\right)-L_{\beta}-1-\pi i\right]\right.
\nonumber \\ &&
\left.+\frac{C_A}{2}\left[-\ln\left(\frac{m^2s}{t_1^2}\right)+L_{\beta}
+\pi i\right]\right\}\, .
\end{eqnarray}
For arbitrary
$\beta$ and fixed scattering angle, we must solve for the relevant  
diagonal basis of color
structure, and determine the eigenvalues.  
At ``absolute" threshold, $\beta=0$, where the heavy quarks are
produced relatively at rest,
we find 
\begin{eqnarray}
\Gamma^{\beta\rightarrow 0}_{S, 11}&=&-\frac{\alpha_s}{\pi}C_F \left(\pi i
+\frac{\pi i}{2 \beta}\right),
\nonumber \\ 
\Gamma^{\beta\rightarrow 0}_{S, 21}&=&0,
\nonumber \\ 
\Gamma^{\beta\rightarrow 0}_{S, 12}&=&\frac{\alpha_s}{2\pi}\left(1-\pi i
-\frac{\pi i}{2 \beta}\right),
\nonumber \\ 
\Gamma^{\beta\rightarrow 0}_{S, 22}&=&\frac{\alpha_s}{\pi}\left[-C_F \left(\pi i
+\frac{\pi i}{2 \beta}\right)
-\frac{C_A}{2}\left(1-\pi i-\frac{\pi i}{2 \beta}\right)\right].
\end{eqnarray}
It is interesting to note  that 
$\Gamma_S^{\beta\rightarrow 0}$ is diagonalized \cite{SotSt} in a basis of singlet and octet
exchange in the $s$ channel, 
\begin{equation}
c_{\rm singlet}=c_1, \quad \quad c_{\rm octet}=-\frac{1}{2N}c_1+\frac{c_2}{2},
\end{equation}
with eigenvalues,
$\lambda^{\beta\rightarrow 0}_{\rm singlet}
=\Gamma^{\beta\rightarrow 0}_{S, 11}$ and 
$\lambda^{\beta\rightarrow 0}_{\rm octet}=\Gamma^{\beta\rightarrow 0}_{S, 22}$.

The general result in this $s$ channel singlet-octet basis
(including the contribution of the self-energy graphs, subtracting
the $\nu^{(i)}$ contribution as noted above and in a general axial gauge) is:
\begin{eqnarray}
\Gamma^{(1,8)}_{S, 11}&=&-\frac{\alpha_s}{\pi}C_F[L_{\beta}+1+\pi i
+\ln(2\sqrt{\nu_a\nu_b})],
\nonumber \\
\Gamma^{(1,8)}_{S, 21}&=&\frac{2\alpha_s}{\pi}
\ln\left(\frac{u_1}{t_1}\right),
\nonumber \\ 
\Gamma^{(1,8)}_{S, 12}&=&\frac{\alpha_s}{\pi}
\frac{C_F}{C_A} \ln\left(\frac{u_1}{t_1}\right),
\nonumber \\
\Gamma^{(1,8)}_{S, 22}&=&\frac{\alpha_s}{\pi}\left\{C_F
\left[4\ln\left(\frac{u_1}{t_1}\right)
-\ln(2\sqrt{\nu_a\nu_b})
-L_{\beta}-1-\pi i\right]\right.
\nonumber \\ &&
\left.+\frac{C_A}{2}\left[-3\ln\left(\frac{u_1}{t_1}\right)
-\ln\left(\frac{m^2s}{t_1u_1}\right)+L_{\beta}+\pi i \right]\right\}\, .
\label{Gammaoneeight}
\end{eqnarray}
$\Gamma_S$ is also diagonalized in this singlet-octet basis
for arbitrary $\beta$ when the parton-parton c.m. scattering angle is
$\theta=90^\circ$ (where $u_1=t_1$) with
eigenvalues that may be read off from Eq.\ (\ref{Gammaoneeight}).

It is of interest, of course, to compare the one-loop expansion of
our results to known one-loop calculations, at the level of NLO.
We give our result as a function of $z$, since the inverse
transforms are trivial.
The result is proportional to the Born cross section,
the one-loop contributions from $g_1^{(q{\bar q})}$ 
and $g_2^{(q{\bar q})}$,
and $\Gamma_{S, 22}^{(1,8)}$.
In the DIS scheme the result is
\begin{eqnarray}
{\hat \sigma}_{f{\bar f}\rightarrow Q{\bar Q}}(1-z,u_1,t_1,s){^{(1)}}
&=&
\sigma_{\rm Born}
\frac{\alpha_s}{\pi} \frac{1}{1-z}\left\{C_F\left[2\ln(1-z)
 +\frac{3}{2} \right.\right.\nonumber \\
&\ &  
\left.\left.+8\ln\left(\frac{u_1}{t_1}\right) 
-2-2 {\rm Re} \, L_{\beta}+2\ln\left(\frac{s}{\mu^2}\right)\right. \right]
\nonumber \\
&\ &  
\left.+C_A\left[-3\ln\left(\frac{u_1}{t_1}\right)+{\rm Re} \, L_{\beta}
-\ln\left(\frac{m^2s}{t_1 u_1}\right)\right]\right\}\, ,
\nonumber\\
\label{longeq}
\end{eqnarray}
where we keep only terms that give rise to $\ln N$.
The logarithm of $s/\mu^2$ describes the evolution of the parton distributions.
This result for the cross section at fixed values of
the $Q{\bar Q}$ invariant mass cannot be compared directly to
the one-loop results of \cite{mengetal},
because  the latter is by contrast a single-particle cross section,
where the singular behavior is given in terms of the variable $s_4$,
\begin{equation}
s_4= (p_2+k)^2-m^2 \approx 2p_2\cdot k\, ,
\end{equation}
where $k=p_a+p_b-p_1-p_2$ is the momentum carried away by the gluon.  At
partonic threshold, both $s_4$ and $(1-z)$ vanish, 
but angular integrals over the gluon momentum reflect the difference
in phase space between fixed $s_4$ and fixed
$1-z \approx 2(p_1+p_2)\cdot k/s$.  

Nevertheless, the cross sections are kinematically equivalent in the 
$\beta\rightarrow 0$ limit, where we may make a direct
comparison.
Near $\beta=0$,   our cross section becomes
\begin{eqnarray}
{\hat \sigma}_{f{\bar f}\rightarrow Q{\bar Q}}(1-z,u_1,t_1,s){^{(1)}}|_{\beta \rightarrow 0}&=&
\sigma_{\rm Born} \frac{\alpha_s}{\pi} \frac{1}{1-z}\left\{C_F\left[2\ln(1-z) 
+\frac{3}{2}\right.\right.
\nonumber \\ &&
\quad\quad \quad \left.\left.
+2\ln\left(\frac{4m^2}{\mu^2}\right)\right]-C_A\right\} \, .
\end{eqnarray}
Near $s=4m^2$, we  may identify $2m^2(1-z)=s_4$, and this expression
becomes identical to the $\beta\rightarrow 0$
limit of eq.\ (30) of \cite{mengetal}.  Even for $\beta>0$, the two cross
sections remain remarkably close, differing only
at first nonleading logarithm in the abelian ($C_F^2$)
term, due to the interplay of angular integrals 
with leading singularities for the differing treatments of
phase space.  

As for the Drell-Yan cross section, our analysis 
applies not only to absolute threshold for the production of the
heavy quarks ($\beta=0$), but also to partonic threshold
for the production of moving heavy quarks.
When $\beta$ nears unity, however,
the anomalous dimensions themselves develop (collinear) singularities
associated with the fragmentation of the heavy quarks, which
in principle may be factored into nonperturbative
fragmentation functions.  

Finally, it is easy to check that the anomalous dimension matrix for heavy
outgoing quarks Eq.\ (\ref{hatGamma1loop}) reduces in the limit $m \rightarrow 0$
to the anomalous dimension matrix for light
outgoing quarks, $q(p_a) + \bar{q}(p_b) \rightarrow \bar{q}(p_1)+q(p_2)$ ,
which is \cite{BottsSt}:
\begin{eqnarray}
\Gamma^{q\bar{q}\rightarrow q\bar{q}}_{S, 11}&=&
\frac{\alpha_s}{\pi}C_F\left[\ln
\left(\frac{v_{ab}v_{12}}{4}\right)
-\frac{1}{2}\ln(\nu_a\nu_b\nu_1\nu_2)+2-2\pi i\right]
\nonumber \\ &&
-\frac{1}{N}\Gamma^{q\bar{q}\rightarrow q\bar{q}}_{S, 21},
\nonumber \\
\Gamma^{q\bar{q}\rightarrow q\bar{q}}_{S, 21}&=&\frac{\alpha_s}{2\pi}
\ln\left(\frac{v_{a2}v_{b1}}{v_{a1}v_{b2}}\right),
\nonumber \\ 
\Gamma^{q\bar{q}\rightarrow q\bar{q}}_{S, 12}&=&\frac{\alpha_s}{2\pi}
\left[\ln\left(\frac{v_{ab}v_{12}}{v_{a1}v_{b2}}\right)-\-2\pi i\right],
\nonumber \\
\Gamma^{q\bar{q}\rightarrow q\bar{q}}_{S, 22}&=&\frac{\alpha_s}{\pi}C_F
\left[\ln\left(\frac{v_{a2}v_{b1}}{4}\right)
-\frac{1}{2}\ln(\nu_a\nu_b\nu_1\nu_2)+2\right]
\nonumber \\ &&
-\frac{1}{N}\Gamma^{q\bar{q}\rightarrow q\bar{q}}_{S, 12}\, .
\end{eqnarray}

\mysection{Applications to $gg \rightarrow Q {\bar Q}$}

In this section we give the results for the anomalous dimension matrix
when the incoming partons are gluons and the outgoing partons are heavy quarks
\begin{equation}
g(p_a)+g(p_b) \rightarrow {\bar Q}(p_1) + Q(p_2)\, .
\end{equation}
For the sake of completeness we also give results for the case
when the outgoing quarks are light
\begin{equation}
g(p_a)+g(p_b) \rightarrow {\bar q}(p_1) + q(p_2)\, .
\end{equation}
In Fig.\ 5 we show the UV divergent one-loop contributions to $S$ for 
$gg \rightarrow Q {\bar Q}$.

\begin{figure}
\centerline{
\psfig{file=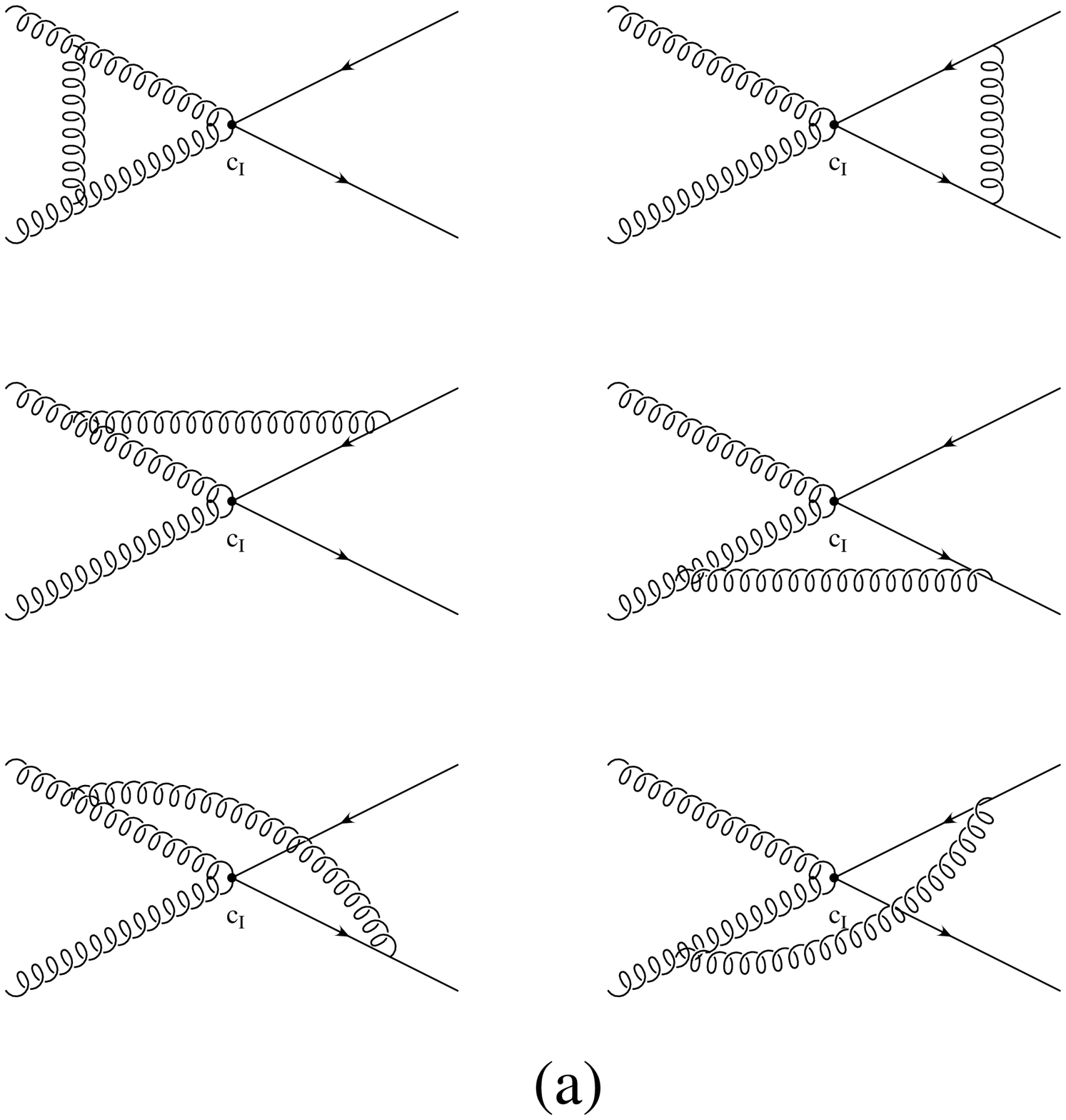,height=4.05in,width=4.05in,clip=}}
\vspace{13mm}
\centerline{
\psfig{file=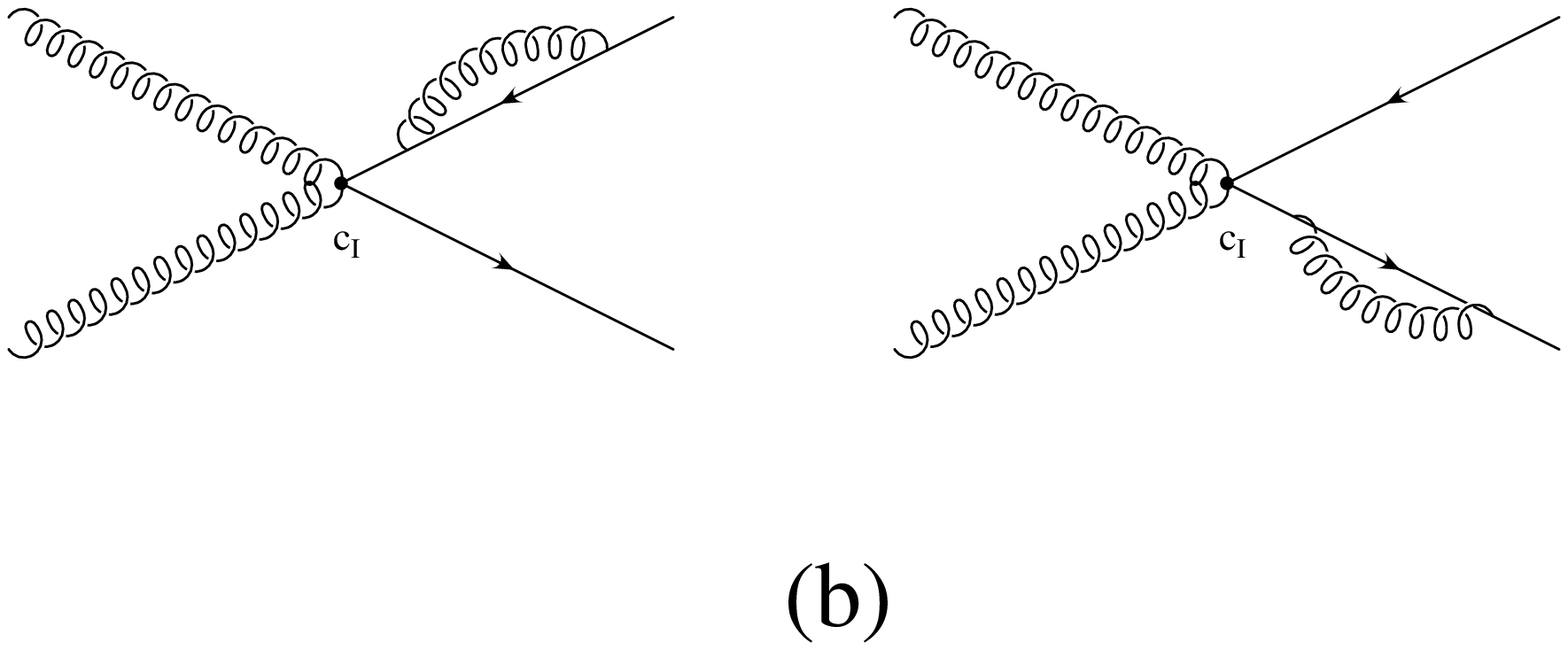,height=1.5in,width=4.05in,clip=}}
{Fig. 5. One-loop corrections to $S_{JI}$ for heavy quark production
through gluon fusion: (a) vertex corrections;
(b) heavy quark self-energy graphs.}
\label{fig. 5}
\end{figure}

Our analysis is similar to the one in the previous section.
We use the same integrals for the calculation of the $\omega^{(I)}$.
The eikonal rules for incoming gluons are slightly modified and are given
in Fig. 6.
We choose the following basis for the color factors:
\begin{equation}
c_1=\delta^{ab}\,\delta_{21}, \quad c_2=d^{abc}\,T^c_{21},
\quad c_3=i f^{abc}\,T^c_{21}.
\end{equation}
Again, the counterterms for $S$ are the ultraviolet divergent
coefficients times our basis color tensors:
\begin{eqnarray}
S_1&=&c_1 Z_{S, 11} + c_2 Z_{S, 21} + c_3 Z_{S, 31},
\nonumber\\
S_2&=&c_1 Z_{S, 12} + c_2 Z_{S, 22} + c_3 Z_{S, 32},
\nonumber\\
S_3&=&c_1 Z_{S, 13} + c_2 Z_{S, 23} + c_3 Z_{S, 33}.
\end{eqnarray}

\begin{figure}
\centerline{\hspace{13mm}
\psfig{file=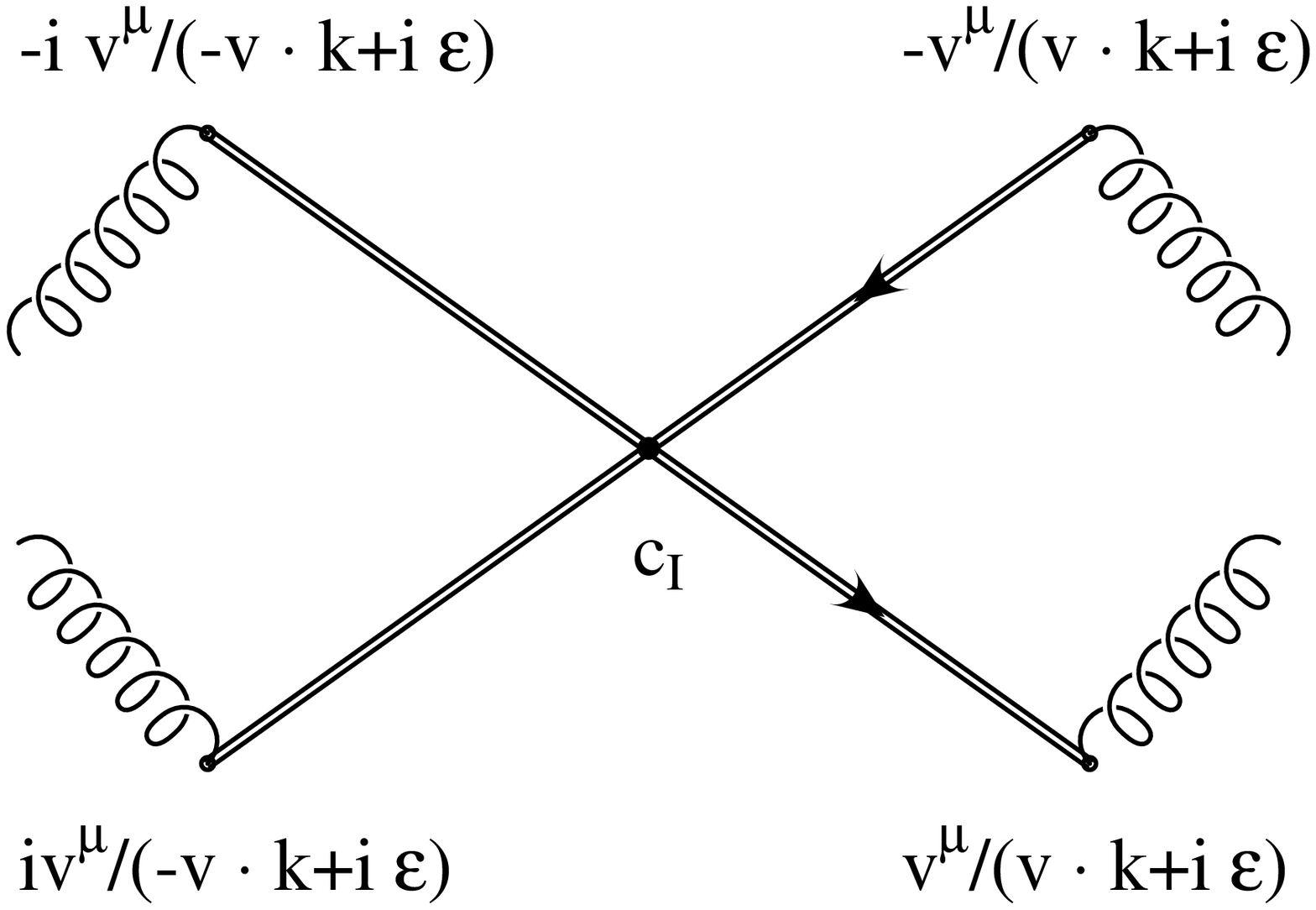,height=2.5in,width=4.05in,clip=}}
{Fig. 6. Feynman rules for eikonal lines in
$gg \rightarrow Q {\bar Q}$. 
The gluon momentum flows out of the eikonal lines. Group
factors for vertices attached to incoming lines are those of three-gluon vertices,
and for outgoing lines those of quark-gluon vertices.}
\label{fig. 6}
\end{figure}

Our results for the  contributions from vertex corrections 
in Fig. 5(a) to the
anomalous dimension matrix when the outgoing quarks are
heavy are:
\begin{eqnarray}
{\hat\Gamma}_{S, 11}&=&\frac{\alpha_s}{\pi}\Bigg\{C_F (-L_{\beta}-L_1-L_2+1)
\nonumber \\
&\ & \quad \quad 
+C_A\left[\ln
\left(\frac{v_{ab}}{2}\right)
-\frac{1}{2}\ln(\nu_a\nu_b)+1-\pi i\right]\Bigg\},
\nonumber \\
{\hat\Gamma}_{S, 21}&=&0,
\nonumber \\ 
{\hat\Gamma}_{S, 31}&=&\frac{\alpha_s}{\pi}
\ln\left(\frac{v_{a2}v_{b1}}{v_{a1}v_{b2}}\right),
\nonumber \\ 
{\hat\Gamma}_{S, 12}&=&0,
\nonumber \\ 
{\hat\Gamma}_{S, 22}&=&\frac{\alpha_s}{\pi}\left\{C_F (-L_{\beta}-L_1-L_2+1)
+\frac{C_A}{2}\left[
\ln\left(\frac{v_{ab}}{2}\right)
\right.\right.
\nonumber \\ &&
\left.\left.
+\frac{1}{2}\ln(v_{a1}v_{b2}v_{a2}v_{b1})+L_{\beta}-\ln\left(\frac{2m^2}{s}\right)
-\ln(\nu_a\nu_b)+2-\pi i\right]\right\},
\nonumber \\ 
{\hat\Gamma}_{S, 32}&=&\frac{N^2-4}{4N}{\hat\Gamma}_{S, 31},
\nonumber \\ 
{\hat\Gamma}_{S, 13}&=&\frac{1}{2}{\hat\Gamma}_{S,31},
\nonumber \\ 
{\hat\Gamma}_{S, 23}&=&\frac{C_A}{4}{\hat\Gamma}_{S,31},
\nonumber \\ 
{\hat\Gamma}_{S, 33}&=&{\hat\Gamma}_{S, 22}.
\label{hatgammaggQQ}
\end{eqnarray}
Again, the gauge dependence associated with the heavy quarks 
is eliminated once we include the 
self-energy graphs in Fig.\ 5(b).
The contribution of the self-energy graphs (in the diagonal
elements only) is, as before, 
\begin{equation}
\frac{\alpha_s}{\pi} C_F(L_1+L_2-2) \; \delta_{IJ}.
\end{equation}
In analogy to the previous section, we subtract $\nu^{(i)}/2$, 
which gives an additional
$-(\alpha_s/\pi)C_A$ shift in the diagonal elements. 
In terms of the Mandelstam invariants $s$, $t_1$, and $u_1$,
the anomalous dimension matrix becomes (in a general axial gauge): 
\begin{eqnarray}
\Gamma_{S,11}&=&\frac{\alpha_s}{\pi}\left [-C_F(L_{\beta}+1)
-C_A\left(\frac{1}{2}\ln\left({4\nu_a\nu_b}\right)+\pi i\right)\right ],
\nonumber \\
\Gamma_{S,21}&=&0,
\nonumber \\
\Gamma_{S,31}&=&\frac{\alpha_s}{\pi}\ln\left(\frac{u_1^2}{t_1^2}\right),
\nonumber \\
\Gamma_{S,12}&=&0,
\nonumber \\
\Gamma_{S,22}&=&\frac{\alpha_s}{\pi}\left\{-C_F(L_{\beta}+1)
-\frac{C_A}{2}\left[\ln\left(4\nu_a\nu_b\right)
-\ln\left(\frac{t_1 u_1}{m^2 s}\right)-L_{\beta}+\pi i
\right]\right\},
\nonumber \\
\Gamma_{S,32}&=&\frac{N^2-4}{4N}\Gamma_{S,31},
\nonumber \\
\Gamma_{S,13}&=&\frac{1}{2}\Gamma_{S,31},
\nonumber \\
\Gamma_{S,23}&=&\frac{C_A}{4}\Gamma_{S,31},
\nonumber \\
\Gamma_{S,33}&=&\Gamma_{S,22}.
\end{eqnarray}

At threshold for the
production of heavy quarks at rest,
the anomalous dimension matrix becomes diagonal.  In
$A^0=0$ gauge (where $\nu_a=\nu_b=1/2)$ its
eigenvalues are
\begin{eqnarray} 
\lambda^{\beta\rightarrow 0}_1&=&\Gamma^{\beta\rightarrow 0}_{S, 11}=\frac{\alpha_s}{\pi}
\left[-C_F\frac{\pi i}{2\beta}-C_A\pi i\right],
\nonumber\\ 
\lambda^{\beta\rightarrow 0}_2&=&\Gamma^{\beta\rightarrow 0}_{S, 22}=\frac{\alpha_s}{\pi}\left[
-C_F\frac{\pi i}{2\beta}
+\frac{C_A}{2}\left(-1-\pi i+\frac{\pi i}{2\beta}\right)\right],
\nonumber\\
\lambda^{\beta\rightarrow 0}_3&=&\Gamma^{\beta\rightarrow 0}_{S, 33}=\lambda^{\beta\rightarrow 0}_2.
\end{eqnarray}
We also note that the matrix is diagonalized at $\theta=90^{\circ}$
for arbitrary $\beta$, with eigenvalues
\begin{eqnarray}
\lambda_1(\theta=90^\circ)&=&\Gamma_{S,11}(\theta=90^\circ)=
{\alpha_s\over \pi}[-C_F (L_{\beta}+1)-C_A\pi i] \, , 
\nonumber\\  
\lambda_2(\theta=90^\circ)&=&\Gamma_{S,22}(\theta=90^\circ)=
{\alpha_s\over \pi}\left\{-C_F (L_{\beta}+1)\right.
\nonumber \\ && \quad \quad
+\left.{C_A\over 2}\left[\ln\left(\frac{t_1^2}{m^2s}\right)+L_{\beta}
-\pi i)\right]\right\}\, ,
\nonumber\\
\lambda_3(\theta=90^\circ)&=&\Gamma_{S,33}(\theta=90^\circ)=
\lambda_2(\theta=90^\circ).
\end{eqnarray}

Again, it is interesting to compare the one-loop expansion of our results
to the one-loop calculations in \cite{mengetal}. 
In this case, our calculation is complicated by the fact
that the color decomposition is not trivial as it was for $q\bar{q}$.
We have to decompose the Born cross section into three terms according to 
our color tensor basis. 
After some algebra our result becomes 
\begin{eqnarray}
{\hat \sigma}_{gg\rightarrow Q{\rm Q}}(1-z,u_1,t_1,s){^{(1)}}&=&
\alpha_s^3 \frac{1}{1-z} K_{gg} B_{\rm QED}(s, t_1, u_1)
\left\{\frac{t_1 u_1}{s^2} \right.
\nonumber \\ &&
\times \left[-8 N^2 C_F C_A 
\left(\ln (1-z)+\frac{1}{2}\ln\left(\frac{s}{\mu^2}\right)\right)\right.
\nonumber \\ &&
-2 N^2 C_F C_A \ln\left(\frac{t_1 u_1}{m^2 s}\right)
\nonumber \\ &&
\left.+4 N^2 C_F \left(C_F-\frac{C_A}{2}\right) {\rm Re} \, L_{\beta}
+4 N^2 C_F^2 \right]
\nonumber \\ &&
+8 N^2 C_F^2 \left(\ln(1-z)
+\frac{1}{2} \ln\left(\frac{s}{\mu^2}\right)\right)
\nonumber \\ &&
+C_F C_A (N^2-2) \ln\left(\frac{t_1 \, u_1}{m^2 s}\right)
\nonumber \\ &&
+ 2 C_F \left(C_F-\frac{C_A}{2}\right) {\rm Re} \, L_{\beta}
\nonumber \\ &&
\left. -2 C_F^2 (N^2-1) \right\} \, ,
\end{eqnarray}
where
\begin{equation}
B_{\rm QED}(s, t_1, u_1)=\frac{t_1}{u_1}+\frac{u_1}{t_1}
+\frac{4m^2s}{t_1 \, u_1}\left(1-\frac{m^2s}{t_1 \, u_1}\right)\, ,
\end{equation}
and $K_{gg}=(N^2-1)^{-2}$ is a color average factor.
The logarithms of $s/\mu^2$ describe the evolution of the parton distributions.

As we discussed in the previous section,
our result cannot be compared directly to the one-loop results
of \cite{mengetal}, but as $\beta\rightarrow 0$ our expression
becomes identical to the $\beta\rightarrow 0$ limit of the sum of
eqs. (36-38) in \cite{mengetal}: 
\begin{eqnarray}
{\hat \sigma}_{gg\rightarrow Q{\bar Q}}(1-z,u_1,t_1,s){^{(1)}}|_{\beta\rightarrow 0}
&=&\alpha_s^3 \frac{1}{1-z} K_{gg}
\left\{4N^2C_F(4C_F-C_A)\right.
\nonumber \\ &&
\times \left[\ln(1-z)+\frac{1}{2}\ln\left(\frac{4m^2}{\mu^2}
\right)\right]
\nonumber \\ &&
\left.+(N^2+2) C_F C_A - 4 N^2 C_F^2 \right\}.
\end{eqnarray}
Again, even for $\beta>0$, the two cross sections
remain remarkably close. 

Finally, the anomalous dimension matrix for the case when 
the outgoing quarks are light is given by
\begin{eqnarray}
\Gamma^{gg\rightarrow q{\bar q}}_{S, 11}&=&\frac{\alpha_s}{\pi}\left\{
C_F\left[\ln\left(\frac{v_{12}}{2}\right)-\frac{1}{2}\ln(\nu_1\nu_2)
+1-\pi i\right]\right.
\nonumber \\ &&
\left.+C_A\left[\ln\left(\frac{v_{ab}}{2}\right)-\frac{1}{2}\ln(\nu_a\nu_b)
+1-\pi i\right]\right\}  ,
\nonumber \\ 
\Gamma^{gg\rightarrow q{\bar q}}_{S, 21}&=&0,
\nonumber \\ 
\Gamma^{gg\rightarrow q{\bar q}}_{S, 31}&=&\frac{\alpha_s}{\pi}
\ln\left(\frac{v_{a2}v_{b1}}{v_{a1}v_{b2}}\right),
\nonumber \\ 
\Gamma^{gg\rightarrow q{\bar q}}_{S, 12}&=&0,
\nonumber \\ 
\Gamma^{gg\rightarrow q{\bar q}}_{S, 22}&=&\frac{\alpha_s}{\pi}\left\{
C_F\left[\ln\left(\frac{v_{12}}{2}\right)-\frac{1}{2}\ln(\nu_1\nu_2)
+1-\pi i\right]\right.
\nonumber \\ && \hskip -0.1 true in
\left.+C_A\left[\frac{1}{4}
\ln(v_{a1}v_{b2}v_{a2}v_{b1})+\frac{1}{2}\ln\left(\frac{v_{ab}}{v_{12}}\right)
-\frac{1}{2}\ln(\nu_a\nu_b)-\ln 2+1\right]\right\} ,
\nonumber \\ 
\Gamma^{gg\rightarrow q{\bar q}}_{S, 32}&=&\frac{N^2-4}{4N}
\Gamma^{gg\rightarrow q{\bar q}}_{S, 31},
\nonumber \\ 
\Gamma^{gg\rightarrow q{\bar q}}_{S, 13}&=&\frac{1}{2}
\Gamma^{gg\rightarrow q{\bar q}}_{S, 31},
\nonumber \\ 
\Gamma^{gg\rightarrow q{\bar q}}_{S, 23}&=&\frac{C_A}{4}
\Gamma^{gg\rightarrow q{\bar q}}_{S, 31},
\nonumber \\ 
\Gamma^{gg\rightarrow q{\bar q}}_{S, 33}&=&\Gamma^{gg\rightarrow q{\bar q}}_{S, 22}.
\label{gglightquark}
\end{eqnarray}
We note that this matrix is also diagonalized at threshold and
at $\theta=90^{\circ}$, and that
the anomalous dimension matrix for heavy
outgoing quarks, Eq.\ (\ref{hatgammaggQQ}) 
reduces to the anomalous dimension matrix for light
outgoing quarks, Eq.\ (\ref{gglightquark}) in the limit $m \rightarrow 0$.

\mysection{Conclusions}
We have illustrated the application of a general method for
resumming next-to-leading logarithms at the edge of phase space
in QCD hard-scattering cross sections.  
We have given explicit results for 
singular distributions at partonic threshold in heavy quark production
through light quark annihilation and gluon fusion, and for light 
quark production through gluon fusion.
Possible extensions include, of course, 
dijet and multijet production, and dijet or heavy
quark transverse momentum distributions.  
Recent estimates of the phenomenological
importance of these nonleading terms to heavy quark production 
indicate that their contribution can be substantial \cite{nkjsrv}.
In addition, an understanding of the all-orders perturbative
cross section in QCD processes may shed light on power corrections
in hadronic scattering \cite{CSt,KoSt,KS,pc}.
We hope that our method will be of use in improving the reliability of 
perturbative QCD calculations for hard scattering.


\mysection*{Acknowledgements}

This work was supported in part by the National Science Foundation,
under grant PHY9309888 and by the PPARC under grant GR/K54601.
We wish to thank 
Lyndon Alvero, Harry Contopanagos, Gregory Korchemsky, Eric Laenen,
and Jack Smith for many helpful conversations.

\end{document}